\documentclass[aps,10pt,prd,notitlepage,nofootinbib]{revtex4-1}

\usepackage[utf8]{inputenc}
\usepackage{newtxtext,newtxmath}
\usepackage{bbold}
\usepackage{bm}
\usepackage{graphicx}
\usepackage{color}
\usepackage{xcolor}
\usepackage{tikz}
\usepackage[colorlinks=true,linkcolor=blue,urlcolor=blue,citecolor=blue]{hyperref}
\usepackage{amsmath,amssymb,amsfonts}
\usepackage{slashed}
\usepackage[english]{babel}
\usepackage{microtype}
\usepackage{dcolumn}
\usepackage{blindtext}
\usepackage{epsfig}
\usepackage{pifont}
\usepackage{dsfont,mathrsfs}
\usepackage{cancel}
\usepackage{bigints}

\usepackage{accents}
\usepackage{soul}
\usepackage{multirow}
\usepackage{appendix}

\newcommand{\nn}{\nonumber}

\newcommand{\ensembleaverage}[1]{\left\langle#1\right\rangle}
\newcommand{\MB}[1]{\left|#1\right|}
\newcommand{\FB}[1]{\left(#1\right)}
\newcommand{\SB}[1]{\left\{#1\right\}}
\newcommand{\TB}[1]{\left[#1\right]}

\newcommand{\scrL}{\mathscr{L}}
\newcommand{\scrM}{\mathscr{M}}

\newcommand{\pp}{p^\prime}
\newcommand{\kp}{k^\prime}

\newcommand{\domg}[1]{ \dfrac{d^3#1}{(2\pi)^3 2E_{#1} }}
\newcommand{\dOmg}[1]{ \dfrac{d^3#1}{(2\pi)^3 E_{#1} }}

\begin{document}
\title{Viscous coefficients and thermal conductivity of a $\pi K N$ gas mixture in the medium}
\author{Pallavi Kalikotay$^{a}$}
\email{orionpallavi@gmail.com}
\author{Nilanjan Chaudhuri$^{b,d}$}
\email{sovon.nilanjan@gmail.com}
\author{Snigdha Ghosh$^{c}$}
\email{snigdha.physics@gmail.com, snigdha.ghosh@saha.ac.in}
\thanks{(Corresponding Author)}
\author{Utsab Gangopadhyaya$^{b}$}
\email{utsabgang@gmail.com}
\author{Sourav Sarkar$^{b,d}$}
\email{sourav@vecc.gov.in}
\affiliation{$^a$Department of Physics, Kazi Nazrul University, Asansol - 713340, West Bengal, India}
\affiliation{$^b$Variable Energy Cyclotron Centre, 1/AF Bidhannagar, Kolkata - 700064, India}
\affiliation{$^c$Saha Institute of Nuclear Physics, 1/AF Bidhannagar, Kolkata - 700064, India}
\affiliation{$^d$Homi Bhabha National Institute, Training School Complex, Anushaktinagar, Mumbai - 400085, India}

\begin{abstract}
The temperature and density dependence of the relaxation times, thermal conductivity, shear viscosity and bulk viscosity for a hot and dense gas consisting of pions, kaons and nucleons have been evaluated in the kinetic theory approach. The in-medium cross-sections for $\pi\pi$, $\pi K$ and $\pi N$ scatterings were obtained by using complete propagators for the exchanged $\rho$, $\sigma$, $K^*$ and $\Delta$ excitations derived using thermal field theoretic techniques. Notable deviations can be observed in the temperature dependence of $\eta$, $\zeta$ and $\lambda$ when compared with corresponding calculations using vacuum cross-sections usually employed in the literature. The value of the specific shear viscosity $\eta/s$ is found to be in agreement with available estimates.
\end{abstract}
\maketitle

\section{INTRODUCTION}
In the past few  decades  ultrarelativistic heavy ion collision (HIC)  experiments  have  been pursued with a lot of vigour essentially because they provide us with the opportunity to examine strongly interacting matter at high energy densities by producing coloured degrees of freedom in a deconfined state known as quark-gluon plasma (QGP) in the  initial stages, followed by a hot hadronic gas mixture~\cite{jadams,Aamodt,Sarkar:2010zza}. Observations at RHIC have presented clear indications that the produced matter behaves more like  a strongly interacting liquid compared to a weakly interacting gas~\cite{Csernai}. This interpretation is primarily due to the fact that the STAR data on elliptic flow of charged hadrons in Au+Au collisions at $ \sqrt{s}=200 $~GeV per nucleon pair could be described (see e.g.~\cite{Mluzum}) using very small but finite values of shear viscosity over entropy density ratio $ \eta / s $  in viscous hydrodynamic simulations.  The estimations of shear $ (\eta) $ and the bulk $ (\zeta) $ viscous coefficients are particularly significant as they can be useful signatures of phase transition or cross-over between quark and hadronic matter. The value of $\eta/s$ shows a minimum near $ T_c $~\cite{Dobado1,Cassing,Csernai,Redlich}, close to the lower bound~\cite{KSS}, whereas $\zeta/s$ shows large or diverging~\cite{Kharzeev,Chen2,Fraile3,Cassing} values around $ T_c $. Thus the effects of dissipation on the dynamical  evolution of matter produced in relativistic heavy ion collisions have been a much discussed topic in recent times~\cite{QMProc}.

Dissipative phenomena are generally studied by considering small deviations from equilibrium  at the microscopic level.  
Transport coefficients  such as shear and bulk viscosity and thermal conductivity are estimated considering the transport of momenta and heat~\cite{Boltzmann_t_eq,zubarev} among the constituents. Quite a few studies on the viscous coefficients have been performed both for quark and hadronic matter~\cite{Gavin,Prakash,Davesne,Santalla,Dobado3,Chen,Itakura,Kharzeev,Fraile1,Dobado1,Chen2,Noronha,Demir,Redlich,Sasaki,Fraile3,Dobado2,Cassing,Rincon,
	Fraile2,Greif,Denicol,Gyulassy,Hosoya,AMY,Arnold2,Rahaman,SabyaPRC89,SabyaPRC90,SabyaBJP}. 
Two different formalisms are generally used for the calculation of transport coefficients of a medium; 
one is the kinetic theory approach~\cite{Landau,Reif,DeGroot:1980dk} where one solves the Boltzmann transport equation, 
the other being the Kubo formalism~\cite{Kubo:1957mj} where one calculates the in-medium spectral functions in the 
long wavelength limit. However, it was demonstrated in Refs.~\cite{Jeon:1994if,Jeon:1995zm} that, in the Kubo formalism, 
a naive skeleton expansion in terms of thermal width may have convergence issues; thus requiring a resummation of all the 
complicated higher-loop diagrams. In contrast the kinetic theory approach being computationally more efficient, has been widely used to evaluate the transport coefficients.

In one of the earlier works~\cite{Gavin} the viscosities and thermal conductivities of both deconfined as well as hadronic matter have been estimated using relaxation time approximation (RTA) in which a variational approach is used to determine the relaxation times. 
Transport coefficients of a single component bosonic system were calculated in~\cite{Davesne} employing Chapman-Enskog method to first order where the  Bose-Einstein distribution function was used instead of the classical one. 
Pions are the most abundant constituents of the hadron gas produced in heavy ion collisions at RHIC and LHC,~\cite{Sarkar:2010zza} which lends justification to the large number of works that can be found in the literature considering a system of only pions in the study of transport coefficients~\cite{Fraile3,Lu,Sukanya_visc,Sukanya_thcon}. Results for  multicomponent hadronic systems, though fewer, have also been reported e.g.~\cite{Prakash,Dobado3,Itakura,Utsab}.
In Ref.~\cite{Santalla}, the shear viscosity of a pion gas is calculated by solving the Uehling-Uhlenbeck equation in which 
the $\pi\pi$ scattering cross section is obtained from chiral perturbation theory. Similar calculations for a multicomponent 
hadronic mixture (composed of $\pi$, $K$ and $\eta$ mesons) were performed in Ref.~\cite{Dobado3}. 
In Ref.~\cite{Itakura}, the authors estimated the shear viscosity ($\eta$) and its ratio to the entropy density ($\eta/s$) of a 
two component hot and dense mixture of pions and nucleons solving the transport equation. 
In Ref.~\cite{Dobado1}, $\eta/s$ calculated in the linear sigma model was found to have a minima near the second order phase transition. 
Similar conclusions regarding the occurrence of maxima of the bulk viscosity to entropy density ratio ($\zeta/s$) near the phase transition has been reported in Refs.~\cite{Chen2,Dobado2}. Analogous behaviour of the viscosities with a minima (maxima) of $\eta/s$ ($\zeta/s$) near 
the phase transition is also reported in Ref.~\cite{Cassing} using parton-hadron-string dynamics (PHSD) off-shell transport approach. 
Calculation of viscosities of a hadron resonance gas (HRG) has been performed in Ref.~\cite{Noronha} and 
an upper bound of the $\eta/s$ near the transition temperature is obtained. 
In Ref.~\cite{Redlich}, transport coefficients of hot quark matter were estimated using the Nambu Jona-Lasinio (NJL) model and their behaviour 
near the chiral phase transition was studied. In addition to these
studies in the kinetic theory approach, the Kubo formalism has been used to study the viscosity of QCD matter in~\cite{Kharzeev} and that of a pion gas in Refs.~\cite{Fraile3,Lang:2012tt}.
In~\cite{SabyaPRC89,SabyaPRC90,SabyaBJP} this formalism has been used to estimate the shear viscous coefficient in a hadronic gas mixture of pions and nucleons. The thermal widths used in these calculations arise from the Landau damping 
of the hadrons in the thermal medium resulting from $2\rightarrow 1$ scattering.

Collisions among constituents are responsible for the transport of momenta, heat etc. within the system and so the scattering cross-section is the principal dynamical input in transport equations where it appears in the collision integral. It is thus necessary to incorporate realistic estimates of cross-sections in interacting hadronic matter.  To this end, Wiranata et al~\cite{Wiranata:2013oaa} have evaluated $\eta/s$ in a hadronic resonance gas with resonances up to mass 2 GeV formed by interactions among the components of a $\pi K N \eta$ mixture. It was shown that the inclusion of more resonances in a multicomponent mixture decreases the value of $\eta$ along with an increase in $s$, both effects serving to decrease $\eta/s$ with increasing temperature.

The hadronic system produced during the later stages of the collision will be at a high temperature and presumably also at finite baryon density keeping in mind the upcoming CBM experiment at FAIR. It is thus pertinent to consider medium effects in the scattering cross-section which is responsible for the dissipative phenomena.  
In this work we consider a gas composed of three components, the pion, kaon and nucleon and intend to demonstrate the effect of the medium on the relaxation times and consequently the thermal conductivity and viscous coefficients of the pion system, pion-kaon system and the pion-kaon-nucleon system in the kinetic theory approach.
We evaluate in particular the in-medium scattering cross-sections for $\pi\pi$, $\pi K$ and $\pi N$ scattering. These interactions are dominated by the $\rho$, $K^*$ and $\Delta$ resonances respectively, whose propagation gets modified in the medium. The effective propagators evaluated using techniques of Thermal Field Theory (TFT) are used in the matrix elements to obtain the in-medium cross-sections of scattering which turn out to be significantly different compared to their vacuum versions usually used in the literature. These were then used in the collision integral. While the left hand side of the transport equations for the different species are treated in the Chapman-Enskog approach the coupled collision integral is handled in the relaxation time approximation.

The article is organized as follows. In the next section we recall the formalism of obtaining transport coefficients using Boltzmann transport equation. In Sec.~\ref{sec.inv.amp} the invariant amplitudes for scattering are discussed followed by the self-energies of the resonances in Sec.~\ref{sec.self.energy}. Sec.~\ref{sec.results} contains numerical results followed by a summary. Some calculational details are  given in the Appendices.


\section{VISCOSITIES AND THERMAL CONDUCTIVITY FROM THE TRANSPORT EQUATION}
In order to obtain the expression for the transport co-efficients we will make use of the relation between thermodynamic forces and the corresponding fluxes given by~\cite{weinberg,DeGroot}
\begin{eqnarray}
&&T^{\mu\nu}=en~u^\mu u^\nu ~-~P\Delta^{\mu\nu}~+~\Pi^{\mu\nu}~+~[(I_q^\mu + h \Delta^{\mu\sigma} N_\sigma) u^\nu ~+~ (I_q^\nu + h \Delta^{\nu\sigma} N_\sigma) u^\mu]~, \\
&&\Pi^{\mu\nu}=2\eta~\langle \partial^\mu u^\nu \rangle ~+~\zeta~(\partial\cdot u)~\Delta^{\mu\nu} \label{viscous_derivative_form}
\end{eqnarray}
where $I_q^\mu$ is the heat flow, $\Pi^{\mu\nu}$  is the viscous part which carries information about the dissipative processes taking place in the system, $P$ is the pressure, $e$ is the energy per particle, $n$ is the particle density, $u^\nu$ is the particle four flow, $\eta$ is the shear viscosity and $\zeta$ is the bulk viscosity. \\

The reduced heat flow $\bar{I}_q^\mu$ can be expressed in terms of heat flow $I^\mu$ as ~\cite{Boltzmann_t_eq}
\begin{eqnarray}
\bar{I}_q^\mu &=& I_q^\mu - \sum_{k} h_k I_k^\mu~,~~~~I_{k}^{\mu}=N_{k}^{\mu}-x_{k}N^{\mu}~, \\
\bar{I}_q^\mu &=& L_{qq} ~\left(\frac{\nabla^\mu T}{T} - \frac{\nabla^\mu P}{nh} \right) + \sum_{j=1}^{N-1} L_{qj} \left( (\nabla^\mu \mu_j)_{P,T} - (\nabla^\mu \mu_N)_{P,T} - \frac{h_j -h_N}{n h} \nabla^\mu P  \right). \label{I_with_Lqq_Lqj}
\end{eqnarray}
Considering mechanical equilibrium i.e. the state with vanishing pressure gradients $\nabla^\mu P=0$ and also considering vanishing gradients of particle fraction i.e.  $\nabla^{\mu}x=0$ we have $(\nabla^{\mu}\mu_{j})_{P,T}=0$, thus Eq.~(\ref{I_with_Lqq_Lqj}) can be written as
\begin{eqnarray}
\bar{I}_q^\mu &=& L_{qq} \frac{\nabla^\mu T}{T}
\end{eqnarray}
where $L_{qq}=\lambda T$, $\lambda$ being the thermal conductivity.

Energy and momentum transfer in the system occurs due to the flow and collision of the constituent particles. In systems out of equilibrium dissipative processes work to bring the system to equilibrium and hence the correlation between transport theory and viscous hydrodynamics is established by taking the distribution function $f_k$ for the $k-$th species to be slightly away from the equilibrium distribution function $f_k^{(0)}$. The measure of the  deviation is given by the quantity $\delta f_k$. Thus we have, 
\begin{equation}
f_k (x,p)=f_k^{(0)}(x,p)+\delta f_k (x,p) \label{dis_f}
\end{equation}
where,
\begin{equation}
\delta f_k(x,p)=f_k^{(0)}(x,p)\left[1 \pm  f_k^{(0)}(x,p)\right]\phi_k (x,p). \label{delta_f}
\end{equation}
The $\pm $ sign in the above expression denotes the Bose enhancement and Pauli blocking. The equilibrium distribution function $f_k^{(0)}$ is given by
\begin{equation}
f_k^{(0)}=\left[\exp\SB{\frac{p\cdot u(x)-\mu_k(x)}{T(x)}}\pm 1\right]^{-1}  
\end{equation}
where $p$ is the four momentum of the particle, $u_\mu (x)$, $\mu_k(x)$ and $T(x)$ are the local flow velocity, chemical potential and temperature respectively. The $\pm$ sign in the distribution function denotes fermions $(+)$ and bosons $(-)$.
In terms of the distribution function, $\Pi^{\mu\nu}$ is given as
\begin{equation}
\Pi^{\mu\nu}=\sum_{k=1}^{N} \int \dOmg{p_k} \Delta^\mu_\sigma \Delta^\nu_\tau p_k^\sigma p_k^\tau \delta f_k \label{Pi_mu_nu}
\end{equation}
where $N$ is the number of different species of particles present in the system and $E_{p_k}=\sqrt{\vec{p}_k^2+m_k^2}$. The reduced heat flow in terms of the distribution function is given by
\begin{eqnarray}
\bar{I}_q^\mu = \sum_{k=1}^N \Delta ^\mu_\sigma ~ g_k \int \frac{d^3p_k}{(2\pi)^3 E_{p_k}} [ (p_k^\nu u_\nu) - h_k] ~p_k^\sigma~ \delta f_k. \label{I_q_mu_int_form}
\end{eqnarray}

In order to realize the form of $\Pi^{\mu\nu}$ as expressed in Eq.~(\ref{viscous_derivative_form}) and $\bar{I}_q^\mu$ as expressed in Eq.~(\ref{I_with_Lqq_Lqj}), the expression for $\phi_k$ which is related to $\delta f_k$ by Eq.~(\ref{delta_f}) is written as a linear combination of thermodynamic forces of different tensorial ranks multiplied by suitable coefficients~\cite{DeGroot}. Thus for the case of thermal conductivity, shear and bulk viscosity $\phi_k$ is chosen to be
\begin{eqnarray}
\phi_k = A_k (\partial\cdot u) -B_{kq}^\nu \Delta_{\mu\nu}  \left(\frac{\nabla^\mu T}{T} \right) - C_k^{\mu\nu}~\langle\partial_\mu u_\nu\rangle \label{phi_k}
\end{eqnarray}

where $A_k$, $B_{kq}^\nu$ and $C_k^{\mu\nu}$ are the unknown coefficients to be determined. The viscous part can be separated into a traceless part and the remainder as	
\begin{equation}
\Pi^{\mu\nu}=\accentset{o}{\Pi}^{\mu\nu} + \Pi \Delta ^{\mu\nu} 
\end{equation}
where $\Pi$	is the viscous pressure
\begin{equation}
\Pi=\sum_{k=1}^{N} \frac{1}{3}  \int \dOmg{p_k} \Delta_{\sigma\tau} p_k^\sigma p_k^\tau \delta f_k~, \label{trace}
\end{equation}
so that
\begin{eqnarray}
\accentset{o}{\Pi}^{\mu\nu}=\Pi^{\mu\nu} -\Pi \Delta^{\mu\nu} 
=\sum_{k=1}^{N} \int \dOmg{p_k} \left\{\Delta^\mu_\sigma \Delta ^\nu_\tau -\frac{1}{3} \Delta_{\sigma\tau}\Delta^{\mu\nu} \right\} p_k^\sigma p_k^\tau \delta f_k. \label{traceless}
\end{eqnarray}

Substituting Eqs.~(\ref{delta_f}) and (\ref{phi_k}) in Eqs.~(\ref{I_with_Lqq_Lqj}), (\ref{trace}) and (\ref{traceless}) and hence making the comparison of the coefficients with Eq.~(\ref{viscous_derivative_form}) the expressions for thermal condcutivity $\lambda$, shear viscosity $\eta$ and bulk viscosity $\zeta$ is obtained as:
\begin{eqnarray}
\lambda &=& - \sum_{k=1}^{N} \frac{1}{3T} g_k \int \frac{d^3p_k}{(2\pi)^3 E_k}~ (p_k^\nu u_\nu - h_k )~f_k^{(0)} (1\pm f_k^{(0)})~ \Delta ^\alpha _\sigma ~p_k^\sigma~ B_\alpha ^{kq},\label{lambda_B} \\
\eta&=&-\sum_{k=1}^{N} \frac{1}{10}\int \dOmg{p_k} \langle {p_k}_\mu {p_k}_\nu \rangle f_k^{(0)} (1\pm f_k^{(0)}) C_k^{\mu\nu}, \label{eta_C} \\
\zeta&=&\sum_{k=1}^{N}\frac{1}{3}\int \dOmg{p_k} \Delta_{\mu\nu} p_k^\mu p_k^\nu f_k^{(0)}(1\pm f_k^{(0)}) A_k. \label{zeta_A}
\end{eqnarray}

To obtain the explicit form of thermal conductivity, shear and bulk viscosity we need to find the unknown coefficients $A_k$, $B_\alpha^{kq}$ and $C_k^{\mu\nu}$ and for that we will make use of the Boltzmann transport equation	
\begin{equation}
p_k^\mu \partial_\mu f_k (x,p)= \sum_{l=1}^{N} \FB{\frac{g_l}{1+\delta_{kl}}}  C_{kl}[f_k] \label{transport_equation}
\end{equation}
where we have considered the binary collision $p_k+p_l\rightarrow p_k'+p_l'$ and $g_l$ is the degeneracy of the $l^\text{th}$ particle. The collision term on the right hand side is
\begin{eqnarray}
C_{kl}[f_k]=\int\int\int \dOmg{p_l} \dOmg{p_k'} \dOmg{p_l'} 
 \Big[f_k(x,p_k')f_l(x,p_l')\SB{1\pm f_k(x,p_k)\frac{}{}} \SB{1\pm f_l(x,p_l) \frac{}{}} \nonumber \\
-f_k(x,p_k)f_l(x,p_l) \SB{1\pm f_k(x,p_k')\frac{}{}}\SB{1\pm f_l(x,p_l') \frac{}{}}  \Big] W_{kl} \label{collision_integral}
\end{eqnarray}
where $W_{kl}$ is the interaction rate. The derivative $\partial_\mu$ can be separated in terms of a temporal and a spatial part in the local rest frame by writing $\partial_\mu= u_\mu D+\nabla_\mu$ where $D=u_\mu \partial^\mu$ and $\nabla_\mu=\Delta_{\mu\nu}\partial^\nu$. Then Eq.~(\ref{transport_equation}) can be written as
\begin{equation}
p^\mu u_\mu Df_k+p^\mu \nabla_\mu f_k=\sum_{l=1}^{N} \FB{\frac{g_l}{1+\delta_{kl}}} C_{kl}[f_k].
\end{equation}

In the Chapman-Enskog approach, the distribution function and its derivative is expanded in terms of $\epsilon$ which is called the non-uniformity parameter or Knudsen number as
\begin{eqnarray}
f&=&f^{(0)}+\epsilon f^{(1)}+\epsilon^2f^{(2)}+ ...~, \nonumber \\
Df &=& \epsilon Df^{(1)} + \epsilon^2 Df^{(2)} + ... ~. \label{d_func_expansion}
\end{eqnarray}
Considering only the  first order  in the above expansion and substituting in Eq.~(\ref{transport_equation}), 
the transport equation reduces to
\begin{equation}
p^\mu u_\mu Df_k^{(0)} + p^\mu \nabla_\mu f_k^{(0)} = \sum_{l=1}^{N} \frac{g_l}{1+\delta_{kl}} C_{kl}[f_k^{(1)}]. \label{transport_1_order}
\end{equation}	
Using conservation equations, the left hand side of the above equation can be simplified to obtain (see Appendix~\ref{sec.app.b})
\begin{equation}
\frac{1}{T} f_k^{(0)}(1\pm f_k^{(0)}) \left[Q_k\partial\cdot u-\langle p_k^\mu~p_k^\nu \rangle\langle \partial_\mu u_\nu\rangle ~+~ (p_k^\sigma u_\sigma -h_k) ~p_k^\mu~ \left( \frac{\nabla_\mu T}{T} \right)  \right]~=~\sum_{l=1}^{N} ~\frac{g_l}{1+\delta_{kl}}~ C_{kl}[f_k^{(1)}]. \label{jump}
\end{equation}	
where
\begin{equation}
Q_k= T^2 \left[-\frac{1}{3}z_k^2 + \tau_k^2\left(\frac{4}{3}-\gamma'\right) +\tau_k\{(\gamma_k''-1)\hat{h}_k -\gamma_k''' \} \right]
\end{equation}
with $z_k=m_k/T$, $\tau_k=(p_k\cdot u)/T$ and $\hat{h}_k=h_k/T$. Here $h_k$ is the enthalpy per particle of type $k$. The steps connecting Eq.~(\ref{transport_1_order}) to Eq.~(\ref{jump})  and the expressions for $\gamma$'s are provided in Appendix~\ref{sec.app.b}.

In order to proceed further and solve Eq.~(\ref{transport_1_order}) we assume that in the interaction 
$p_k+p_l\rightarrow p_k' +p_l'$, only particles with momentum $p_k$ are out of equilibrium and the remaining ones 
i.e. particles with momentum $p_l$, $p_k'$ and $p_l'$ are in equilibrium. This assumption is the well known 
Relaxation Time Approximation (RTA). Thus substituting $f_k^{(1)}$, $f_l^{(1)}$, $f_k^{(1')}$ and $f_l^{(1')}$ 
in Eq.~(\ref{collision_integral}) with $f_k^{(0)}+\delta f_k$, $f_l^{(0)}$, $f_k^{(0')}$ and $f_l^{(0')}$ 
respectively, the rhs of Eq.~(\ref{transport_1_order}) reduces to
\begin{equation}
\sum_{l=1}^{N}\FB{\frac{g_l}{1+\delta_{kl}}} C_{kl}[f_k] = -\frac{\delta f_k}{\tau_k} E_k
\label{rta_app}
\end{equation}	
where,
\begin{equation}
\tau_k= \sum_{l=1}^{N} \FB{\frac{1}{\tau_{kl}^{-1}}}
\label{relax}
\end{equation}
with 
\begin{eqnarray}
\TB{\tau_{kl}}^{-1} &=& \FB{\frac{g_l}{1 + \delta_{kl}}} \dfrac{1}{2 E_{p_k}} \int\int\int \domg{p_l} \domg{p_{k'}} \domg{p_{l'}}(2\pi)^4 \delta ^4 \FB{p_k + p_l -p_{k'} -p_{l'} }\MB{ \scrM_{kl} }^2  \nn \\  
&&  \hspace{1.5in } \times \TB{\frac{}{}f_l^{(0)}(1\pm f_{k'}^{(0)})(1\pm f_{l'}^{(0)})\mp (1\pm f_l^{(0)})f_{k'}^{(0)}f_{l'}^{(0)}} \nn \\ 
&=& \FB{\frac{g_l}{1 + \delta_{kl}}} \dfrac{{\rm csh }(\epsilon_k/2)}{E_k} \int\int\int d\Gamma_{p_l} d\Gamma_{p_{k'}} d\Gamma_{p_{l'} } (2\pi)^4 \delta ^4 \FB{p_k + p_l -p_{k'} -p_{l'} } \MB{ \scrM_{kl} }^2  \label{tau_ab}
\end{eqnarray}

in which, for four-momenta $ q $, $ d\Gamma_q = \dfrac{1}{2{\rm csh }(\epsilon_k/2)} \dfrac{d^3q}{(2\pi)^2 2E_{q} }$, $\epsilon_q = \dfrac{E_q -\mu_q}{T} $ and ${\rm csh }(\epsilon_q) = \cosh( \epsilon_q) \FB{\dfrac{}{}\sinh(\epsilon_q)} $ if $q$ represents a 
Fermion(Boson). In the above equations, $\scrM_{kl}$ is the invariant amplitude for elastic $kl\rightarrow kl$ scattering. 
Following Ref.~\cite{Prakash}, we will assume $ f_{p'}^{(0)} \sim f_p^{(0)} $ and $ f_{k'}^{(0)} \sim f_k^{(0)} $ so that we can analytically integrate over the momenta of final particles $\kp$ and $\pp$ respectively. 
This results in the expression for the relaxation time as
\begin{equation}
\TB{\tau_k}^{-1} = \sum_{l=1}^{N} \FB{\frac{g_l}{1 + \delta_{kl}}} \int \dfrac{d^3 p_{k}}{(2\pi)^3} \FB{\sigma^{kl} v_{\rm rel}^{kl}} 
f_l\FB{1 \pm f_l}. \label{tau_k}
\end{equation}

Thus, using the RTA and Eq.~(\ref{delta_f}), the transport equation can be reduced to
\begin{equation}
\frac{1}{T E_k}\left[Q_k~\partial\cdot u~-~\langle p_k^\mu~p_k^\nu\rangle \langle\partial_\mu~u_\nu\rangle~+~ (p_k^\sigma u_\sigma -h_k)~ p_k^\mu~ \left( \frac{\nabla_\mu T}{T} \right)\right]=~-\frac{\phi_k}{\tau_k}. \label{transport_RTA}
\end{equation}
Substituting the expression for $\phi_k$ from Eq.~(\ref{phi_k}) in Eq.~(\ref{transport_RTA})  and comparing the coefficients of $(\partial \cdot u)$, $\langle \partial_\mu u_\nu\rangle $ and $\frac{\nabla_\mu T}{T}$ on both sides the unknown quantities $A_k$, $B_{kq}^\nu$ and $C_k^{\mu\nu}$ is found to be:
\begin{eqnarray}
A_k&=&-\frac{\tau_k}{T E_k} Q_k~, \label{A} \\
B_\nu^{kq}~\Delta^{\mu\nu}&=& \frac{\tau_k}{E_k T} ~ (p_k^\sigma u_\sigma - h_k)~ \Delta^{\mu\nu}~ p_{k\nu}~, \label{B} \\
C_k^{\mu\nu} &=& -\frac{\tau_k}{T E_k} \langle p_k^\mu  p_k^\nu\rangle. \label{C}
\end{eqnarray}

Finally, substituting the values of $A_k$, $B_{kq}^\nu$ and $C_k^{\mu\nu}$ in Eq.~(\ref{lambda_B}), (\ref{zeta_A}) and (\ref{eta_C}) respectively we obtain the final expressions for thermal conductivity, shear and bulk viscosities as

\begin{eqnarray}
\lambda &=&\frac{1}{3T^2} \sum_{k=1}^{N} \int~\frac{d^3 p_k}{(2\pi)^3}~\frac{g_k \tau_k}{E_{p_k}^2}~ p_k^2 ~  (p_k^\nu u_\nu -h_k)^2 ~ f_k^{(0)}(1\pm f_k^{(0)}), \label{lambda} \\
\eta &=& \frac{1}{15 T}\sum_{k=1}^{N} \int\frac{d^3 p_k}{(2\pi)^3}\frac{g_k \tau_k}{E_{p_k}^2}|\vec{p}_k|^4
f_k^{(0)}(1\pm f_k^{(0)}), \label{eta} \\
\zeta &=&\frac{1}{T}\sum_{k=1}^{N}\int \frac{d^3 p_k}{(2\pi)^3}\frac{g_k \tau_k}{E_{p_k}^2}Q_k^2f_k^{(0)} (1\pm f_k^{(0)}). \label{zeta}
\end{eqnarray}

It is clear from the above expressions that the relaxation times are the essential dynamical components responsible 
for dissipative processes occurring in the system evolving towards equilibrium. Evaluation of the relaxation times for 
the pion, kaon and nucleon in the medium using effective interactions constitutes the main part of this work and is discussed in the next section. 

Before we end this section a few comments on the use of the relaxation time approximation (RTA) are in order. This simplistic approach of treating the collision integral is limited by the fact that it is not possible to have control over the degree of accuracy of the method and neither can one go to higher orders to increase the accuracy as is possible in the Chapman-Enskog (CE) approach. In addition, the latter method involves the transport cross-section with an angular weight of $(1-\cos^2\theta)$ in first-order calculations which accounts for the momentum transfer in collisions desirable for the evaluation of viscosities is lacking in the RTA featuring the total cross-section. These aspects have been discussed in~\cite{Wiranata:2012br} and~\cite{Plumari:2012ep} along with a comparison of the two methods for various cases. The study by Wiranata et al~\cite{Wiranata:2012br} reveal that the extent of agreement  between the CE and the RTA approaches depends sensitively on the energy dependence of the differential cross sections employed. It was shown that for an interacting pion gas where the cross-section involves the $\rho$ resonance, the ratio of shear viscosities calculated in the CE and RTA methods varies between 1.18 at a temperature of 100 MeV to about 1.1 at 160 MeV. 
Since the cross-sections used here have a similar nature as far as the energy dependence is concerned, this not so large disagreement could justify use of the simplistic approach of RTA keeping in mind that the present study is basically aimed at  bringing out the relative effect brought about by the in-medium cross-sections on the temperature dependence of the transport coefficients compared to the vacuum ones.

\section{INVARIANT AMPLITUDES} \label{sec.inv.amp}
The in-medium cross-sections are obtained in the following manner. 
In the matrix elements for $2\rightarrow2$ scattering processes, evaluated from well-known effective interaction Lagrangians, 
the vacuum propagators corresponding to the intermediate resonances appearing in $s$-channel diagrams, are replaced with 
effective ones obtained from a Dyson-Schwinger sum containing one-loop self-energy diagrams in {\it vacuum}. 
This introduces an imaginary part in the matrix elements rendering a Breit-Wigner like structure to the cross-section.  
They are normalized to experimental data fixing a few unknown model parameters in the process. 
The corresponding in-medium cross-sections are then obtained by evaluating the self-energy diagrams in the {\it medium }
using standard techniques of finite temperature field theory.

Let us begin by evaluating the matrix elements of $\pi(k)\pi(p)\rightarrow\pi(k')\pi(p')$ 
scattering using the effective Lagrangian for $\rho\pi\pi$ and $\sigma\pi\pi$ 
interactions~\cite{Serot:1984ey}
\begin{eqnarray}
\scrL_\text{int} = g_{\rho\pi\pi}\vec{\rho}_\mu\cdot\vec{\pi}\times\partial^\mu\vec{\pi} + 
\frac{1}{2}g_{\sigma\pi\pi}m_\sigma\vec{\pi}\cdot\vec{\pi}\sigma~. \label{eq.lagrangian}
\end{eqnarray}
The values of the coupling constants in Eq.~(\ref{eq.lagrangian}) follow from the 
experimental decay widths of $\rho$ and $\sigma$ mesons and we get $g_{\rho\pi\pi}=6.05$ and 
$g_{\sigma\pi\pi}=2.5$. It is convenient to use the isospin basis for expressing the 
invariant amplitudes in different isospin channels. Denoting the invariant amplitude in 
a channel with total isospin $I$ by $\scrM^{\pi\pi}_I$, we get~\cite{Sukanya_1}
\begin{eqnarray}
\scrM^{\pi\pi}_2 &=& g_{\rho\pi\pi}^2\left[-\left(\frac{s-u}{t-m_\rho^2}\right)-
\left(\frac{s-t}{u-m_\rho^2}\right) \right] +4g_{\sigma\pi\pi}^2\left[\frac{1}{t-m_\sigma^2} + \frac{1}{u- m_\sigma^2}\right], \\
\scrM^{\pi\pi}_1 &=& g_{\rho\pi\pi}^2\left[ 2\left(\frac{t-u}{s-m_\rho^2 - \Pi_\rho}\right)+\left(\frac{s-u}{t-m_\rho^2}\right)
-\left(\frac{s-t}{u-m_\rho^2}\right)\right] + 4g_{\sigma\pi\pi}^2\left[\frac{1}{t-m_\sigma ^2}- \frac{1}{u - m_\sigma^2}\right], \\
\scrM^{\pi\pi}_0 &=& g_{\rho\pi\pi}^2\left[2\left(\frac{s-u}{t- m_\rho^2}\right)+2\left(\frac{s-t}{u-m_\rho^2}\right)\right]
+4g_{\sigma\pi\pi}^2\left[\frac{3}{s-m_\sigma^2 - \Pi_\sigma} + \frac{1}{t- m_\sigma^2} + \frac{1}{u- m_\sigma^2}\right]
\end{eqnarray}
where $s=(k+p)^2$, $t=(k-k')^2$ and $u=(k-p')^2$ are the Mandelstam variables. It is to be noted 
that, we have replaced the vacuum $\rho$ and $\sigma$ propagator in the $s$-channel diagrams 
by the complete ones obtained from a Dyson-Schwinger sum involving the one-loop self energies of 
$\rho$ and $\sigma$ mesons denoted by $\Pi_\rho$ and $\Pi_\sigma$ respectively. The calculations 
of the self energies will be discussed in Sec.~\ref{sec.self.energy}.

The calculation of the invariant amplitudes for $\pi(k)N(p)\rightarrow\pi(k')N(p')$,  
$\pi(k)K(p)\rightarrow\pi(k')K(p')$ and $K(k)K(p)\rightarrow K(k')K(p')$ 
are done in a similar way. In this case the interaction 
Lagrangians read~\cite{Krehl,Ko:1993id}
\begin{eqnarray}
\scrL_{\pi N\Delta} &=& \frac{f_{\pi N\Delta}}{m_\pi}\bar{\Delta}_\alpha{\cal O}^{\alpha\mu}
\vec{\bm{T}}^\dagger\cdot\partial_\mu\vec{\pi}\Psi + \text{Hermitian Conjugate} \label{eq.Lagrangian.piN}\\
\scrL_{\pi KK^*} &=& i g_{\pi KK^*} \overline{K}^*_\mu\vec{\bm{\tau}}\cdot\left[ K \FB{\partial^\mu \vec{\pi}} - 
\FB{\partial^\mu K} \vec{\pi}\frac{}{}\right] + \text{Hermitian Conjugate} \label{eq.Lagrangian.piK}\\
\scrL_{KK\phi} &=& i g_{KK\phi}\left[\overline{K}(\partial_\mu K) - (\partial_\mu \overline{K}) K \right] \phi^\mu~.
\end{eqnarray}
where, $\Delta^\mu=\left[\begin{array}{c} \Delta^{++} \\ \Delta^{+} \\ \Delta^{0} \\ \Delta^{-} \end{array} \right]^\mu$ is the $\Delta$ isospin quadruplet, 
$\Psi=\left[\begin{array}{c} p \\ n \end{array} \right]$,   
$K=\left[\begin{array}{c} K^+ \\ K^0 \end{array} \right]$ and 
$K^*_\mu=\left[\begin{array}{c} K^{*+} \\ K^{*0} \end{array} \right]_\mu$ are respectively the isospin doublets for 
the nucleon, Kaon and $K^*$. The coupling constants are analogously fixed from the experimental decay widths 
of $\Delta$ and $K^*$ and we get $f_{\pi N\Delta}=2.8$, $g_{\pi KK^*}=10.80$. 
The invariant amplitudes in different isospin channels are given by
\begin{eqnarray}
\MB{\scrM^{\pi N}_{3/2}}^2&=&\frac{1}{2}\left(\frac{f_{\pi N\Delta}}{m_\pi}\right)^4
\left[ \frac{T_s}{\left|s-m_\Delta^2-\Pi_\Delta\right|^2} +
\frac{T_u}{\left(u-m_\Delta^2\right)^2} + \frac{2T_m (s-m_\Delta^2-\text{Re}{\Pi_\Delta})}
{3(u-m_\Delta^2)\left|s-m_\Delta^2-\Pi_\Delta \right|^2} \right], \label{eq.modm.squared.pin.32} \\
\MB{\scrM^{\pi N}_{1/2}}^2&=&\frac{1}{2}\left(\frac{f_{\pi N\Delta}}{m_\pi}\right)^4
\FB{\frac{16}{9}}\TB{\frac{T_u}{\FB{u-m_\Delta^2}^2}}, \label{eq.modm.squared.pin.12} \\
\scrM^{\pi K}_{3/2} &=& 2g_{\pi KK^*}^2 \left[\frac{(t-s) + (m_k^2 - m_\pi^2)/m_{K^*}^2}{u-m_{K^*}^2 } \right],  \\
\scrM^{\pi K}_{1/2} &=& g_{\pi KK^*}^2 \left[3\frac{(t-u)+(m_k^2 - m_\pi^2)/m_{K^*}^2}{s- m_{K^*}^2 - \Pi_{K^*}} -\frac{(t-s) + (m_k^2 - m_\pi^2)/m_{K^*}^2}{u-m_{K^*}^2 } \right], \\
\scrM^{KK}_1 &=& g_{KK\phi}^2\left[\frac{u-s}{t- m_\phi^2} + \frac{t-s}{u-m_\phi^2}\right], \\
\scrM^{KK}_0 &=& g_{KK\phi}^2\left[\frac{u-s}{t- m_\phi^2} - \frac{t-s}{u-m_\phi^2}\right]
\end{eqnarray}
where, $T_s$, $T_u$ and $T_m$ contain traces over Dirac matrices and details can be found in Ref.~\cite{Snigdha}. 
In the above equations, $\Pi_\Delta$ and $\Pi_{K^*}$ are the one-loop self energies of $\Delta$ and $K^*$ 
which will be obtained in the next section. It is worth mentioning that, to take into account the finite 
size effect of the hadrons we have considered hadronic form factors 
$F(p,k) =\Lambda^2\TB{\Lambda^2+\frac{(p\cdot k)^2}{m_p^2} -k^2}^{-1}$ in each of 
the $\pi N\Delta$ and $\pi KK^*$ vertices where $p$ is the momentum of nucleon/kaon and $k$ is the momentum 
of pion. In this work we have taken $\Lambda_{\pi N}=600$ MeV and $\Lambda_{\pi K}=350$ MeV. In Eqs.~(\ref{eq.modm.squared.pin.32}) and (\ref{eq.modm.squared.pin.12}), we have made an average over initial spin states and a sum over final spin states of the nucleon.

Since we will be calculating isospin averaged cross sections, we define the corresponding isospin averaged 
invariant amplitude by
\begin{eqnarray}
\overline{\MB{\scrM}^2} &=& \sum_{I}(2I + 1)\MB{\scrM_I}^2 \Big/ \sum_{I} (2I+1)
\end{eqnarray}
which is used to obtain the cross-section from 
\begin{eqnarray}
\sigma(s)=\frac{1}{64\pi^2 s}\int\ d\Omega\overline{\MB{\scrM}^2}.
\end{eqnarray}


\section{ONE-LOOP SELF ENERGIES OF $\rho$, $\sigma$, $\Delta$ AND $K^*$} \label{sec.self.energy}

The one-loop self energies $\Pi_h(q)$ of different hadrons $h\in\{\rho,\sigma,\Delta,K^*\}$ 
at finite temperature and density can be calculated using the standard techniques 
of Real Time Formalism (RTF) of TFT~\cite{HadronBook,bellac}. Contributions to $\Pi_h$ come from different loop graphs containing 
other hadrons ($i,j$). In a most general notation, the real part of the self energy reads
\begin{eqnarray}
\text{Re}~\Pi_h(q) &=& \sum_{ \{i,j\} \in \{\text{loops}\}}  
\int\frac{d^3k_i}{(2\pi)^3} \frac{1}{2 \omega_{k_i}\omega_{p_j}}\mathcal{P}\left[
\left(\frac{n_+^{k_i}\omega_{p_j} \mathcal{N}_{ij}^h(k_i^0=\omega_{k_i})}{(q_0-\omega_{k_i})^2-\omega_{p_j}^2}\right) +
\left(\frac{n_-^{k_i}\omega_{p_j} \mathcal{N}_{ij}^h(k_i^0=-\omega_{k_i})}{(q_0+\omega_{k_i})^2-\omega_{p_j}^2}\right)\right. \nn\\ 
&&\hspace{3cm}\left. -\left(\frac{a_jn_+^{p_j}\omega_{k_i} \mathcal{N}_{ij}^h (k_i^0=q_0-\omega_{p_j})}{(q_0-\omega_{p_j})^2-\omega_{k_i}^2}\right) -
\left(\frac{a_jn_-^{p_j}\omega_{k_i} \mathcal{N}_{ij}^h (k_i^0=q_0+\omega_{p_j})}{(q_0+\omega_{p_j})^2-\omega_{k_i}^2}\right) \right] \label{eq.repi}
\end{eqnarray}
whereas the imaginary part is
\begin{eqnarray}
\text{Im}~\Pi_h(q)&=& -\pi\epsilon(q_0) \sum_{ \{i,j\} \in \{\text{loops}\}} 
\int\frac{d^3k_i}{(2\pi)^3}\frac{1}{4 \omega_{k_i}\omega_{p_j}}\nn \\
&&\times \left[\frac{}{}\mathcal{N}_{ij}^h(k_i^0=\omega_{k_i})\left\{(1+n_+^{k_i}-a_jn_+^{p_j})
\delta(q_0-\omega_{k_i}-\omega_{p_j})+(-n_+^{k_i}-a_jn_-^{p_j})
\delta(q_0-\omega_{k_i}+\omega_{p_j})\right\} \right. \nn\\ 
&& \left.+ \mathcal{N}_{ij}^h(k_i^0=-\omega_{k_i})\left\{(-1-n_-^{k_i}+a_jn_-^{p_j})
\delta(q_0+\omega_{k_i}+\omega_{p_j})+
(n_-^{k_i}+a_jn_+^{p_j})\delta(q_0+\omega_{k_i}-\omega_{p_j})\right\}\frac{}{}\right]~.
\label{eq.impi}
\end{eqnarray}
where, the distribution functions for loop particles are given by 
$n_\pm^{k_i} = \left[ e^{\beta\left(\omega_{k_i}\mp \mu_i \right)}\mp 1 \right]^{-1}$ (according to boson/fermion) 
with $\omega_{k_i} = \sqrt{\vec{k}_i^2+m_i^2}$. In the above equations, $i$-type particles are always 
bosons and $a_j=\mp1$ depending upon whether the $j$-type particle is a boson/fermion. The self energy may 
contain additional Lorentz/Dirac indices, however in this work we have used the spin/polarization averaged self 
energies for the calculation of the $S$-matrix elements. It is worth 
mentioning that, if the $i$ and/or $j$ type particles are unstable, then the self energies are folded with 
the vacuum spectral functions of the loop particles.

The $\rho$ self energy consists of $\{i,j\}$ = $\{\pi,\pi\}$, $\{\pi,\omega\}$, $\{\pi,h_1\}$ and $\{\pi,a_1\}$ 
loops whereas the $\sigma$ self energy has contribution from only $\{i,j\}$ = $\{\pi,\pi\}$ loop. 
The detailed expressions of $\mathcal{N}_{ij}^\rho$ and $\mathcal{N}_{ij}^\sigma$ can be read from 
Ref.~\cite{Sukanya_1}. 
For $\Delta$ self energy, we consider loop graphs containing 
$\{i,j\}$ = $\{\pi,N\}$, $\{\rho,N\}$, $\{\pi,\Delta\}$ and $\{\rho,\Delta\}$ and detailed  
expressions of $\mathcal{N}_{ij}^\Delta$ may be found in Ref.~\cite{Snigdha}. For the $K^*$, the contribution to 
the self energy comes from $\{i,j\}$ = $\{\pi,K\}$ and $\mathcal{N}_{\pi K}^{K^*}$ is given by
\begin{equation}
\mathcal{N}_{\pi K}^{K^*}=g_{\pi KK^*}^2\left[-(q-2k)^2+\frac{1}{m_{K^*}^2}(q^2-2q\cdot k)^2 \right].
\end{equation}

The four different terms containing the Dirac delta functions in imaginary part of the self energy 
correspond to different physical processes like decay and scattering owing to the annihilation of hadron $h$ in the thermal medium.

\section{NUMERICAL RESULTS} \label{sec.results}

\begin{figure}[h]
	\begin{center}
		\includegraphics[angle=-90, scale=0.35]{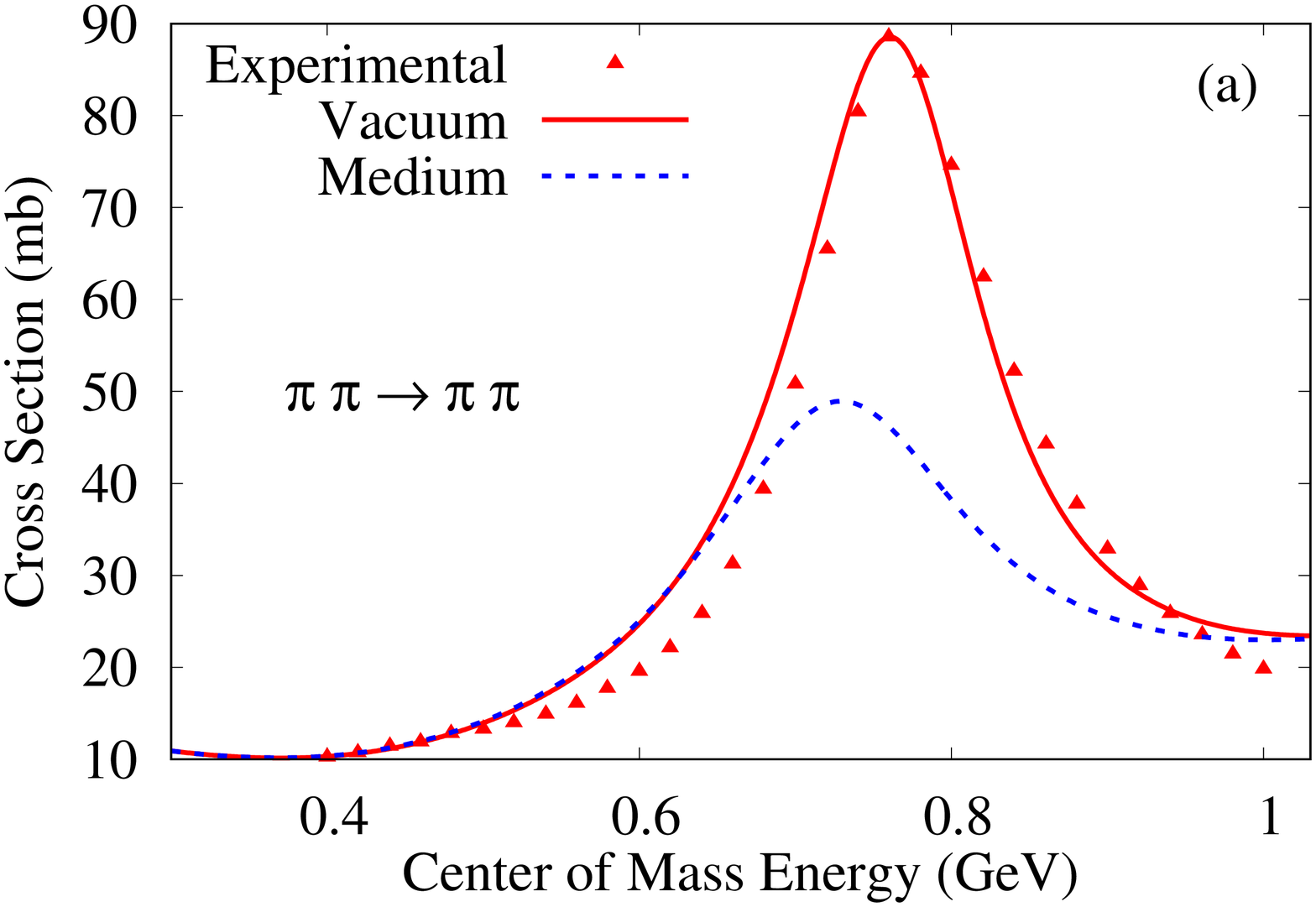}
		\includegraphics[angle=-90, scale=0.35]{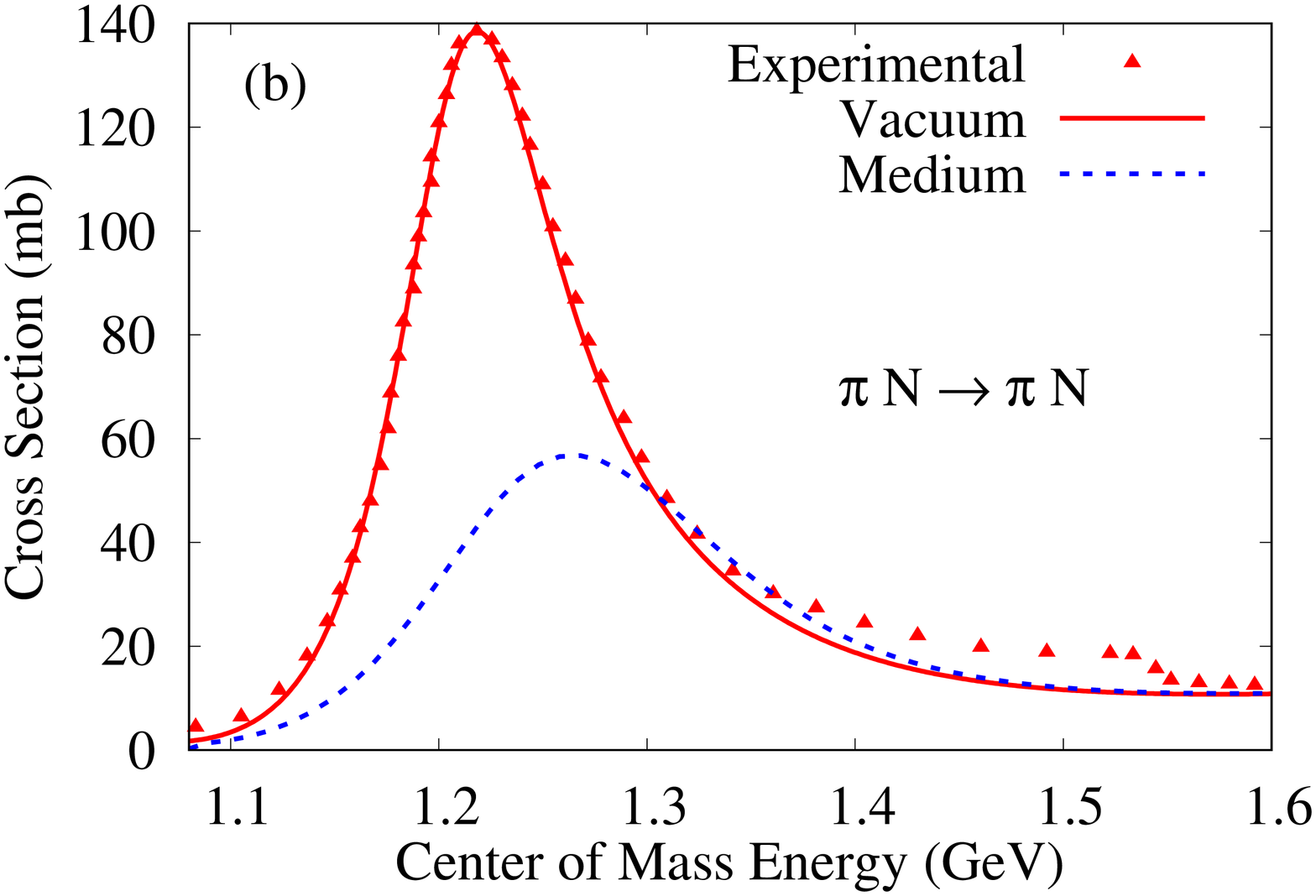} 
		\includegraphics[angle=-90, scale=0.35]{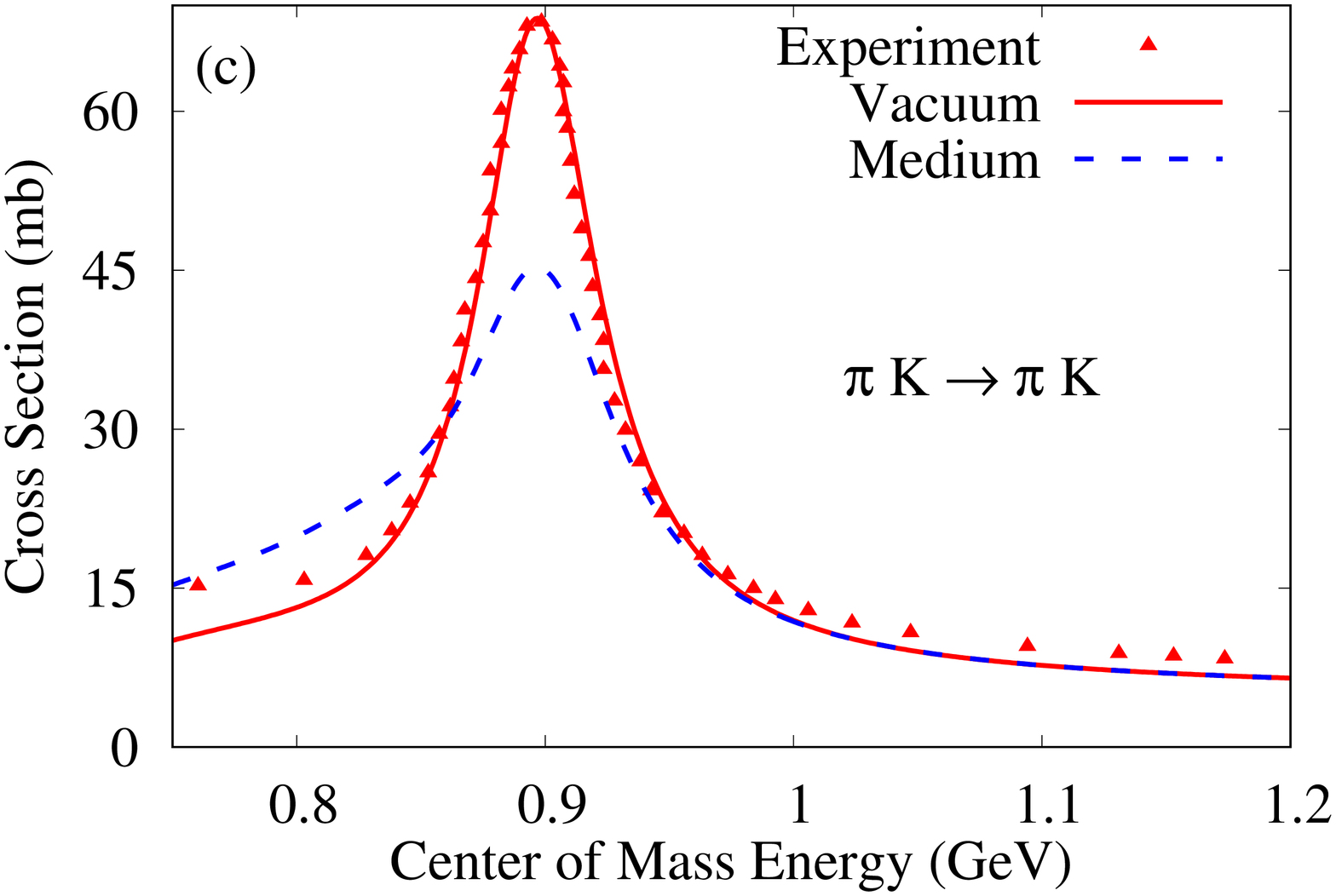}
	\end{center}
	\caption{The (a)$ \pi \pi \rightarrow\pi \pi $, (b)$ \pi N\rightarrow\pi N $ and (c) $ \pi K \rightarrow \pi K $ elastic scattering cross section as a function of centre of	mass energy compared among experiment, vacuum and medium corresponds to $ T = 160 $ MeV, $\mu_\pi=\mu_{K}=0$ and $ \mu_N=200 $ MeV. Experimental data have been taken from Ref.~\cite{Prakash} }
	\label{cross_sec}
\end{figure}

We begin this section by discussing the elastic scattering cross sections for $\pi\pi\rightarrow\pi\pi$, 
$\pi N\rightarrow\pi N$ and $\pi K\rightarrow\pi K$. In Fig.~\ref{cross_sec} both vacuum and in-medium cross-sections are plotted along with the experimental data~\cite{Prakash}. 
Since the matrix elements obtained in Sec.~\ref{sec.inv.amp} contain the one-loop in-medium self energies of $\rho$, $\sigma$, 
$\Delta$ and $K^*$, the scattering cross sections also depend on the temperature and density of the thermal medium. 
Using the approach described above we have been able to obtain a very good fit of the vacuum cross-section with 
the experimental data for the given set of model parameters for the tree types of scattering mentioned above. Having thus fixed our model in vacuum we replace the propagators with their thermal versions as described above to obtain the in-medium cross-section. The broadening of the widths of the resonances in the medium are reflected in the suppression of the cross section at the resonance energy and this is seen to be about $ 50-70 \% $  at $ T=160$ MeV. 
The small shift lateral of the peak of the cross section is due to the small contribution from the real part of the 
thermal self energy function.
\begin{table}[h]
	\begin{center}
		\begin{tabular}{cccc}
			\hline
			Chemical potential  & ~~~~~~~ $\mu_\pi$ ~~~~~~~ & ~~~~~~~ $\mu_K$ ~~~~~~~ & ~~~~~~~ $\mu_N$ ~~~~~~~ \\
			\hline 	\hline 
			Set-1 & 0 & 0 & 0\\ 
			Set-2 & 0 & 0 & 200\\ 
			Set-3 & 50 & 100 & 200 \\
			\hline	
		\end{tabular}
	\end{center}
	\caption{ Different set of values of $\pi$, $K$ and $N$ chemical potentials used in this work. }
	\label{table1}
\end{table}

\begin{figure}[h]
	\begin{center}
		\includegraphics[angle=-90, scale=0.35]{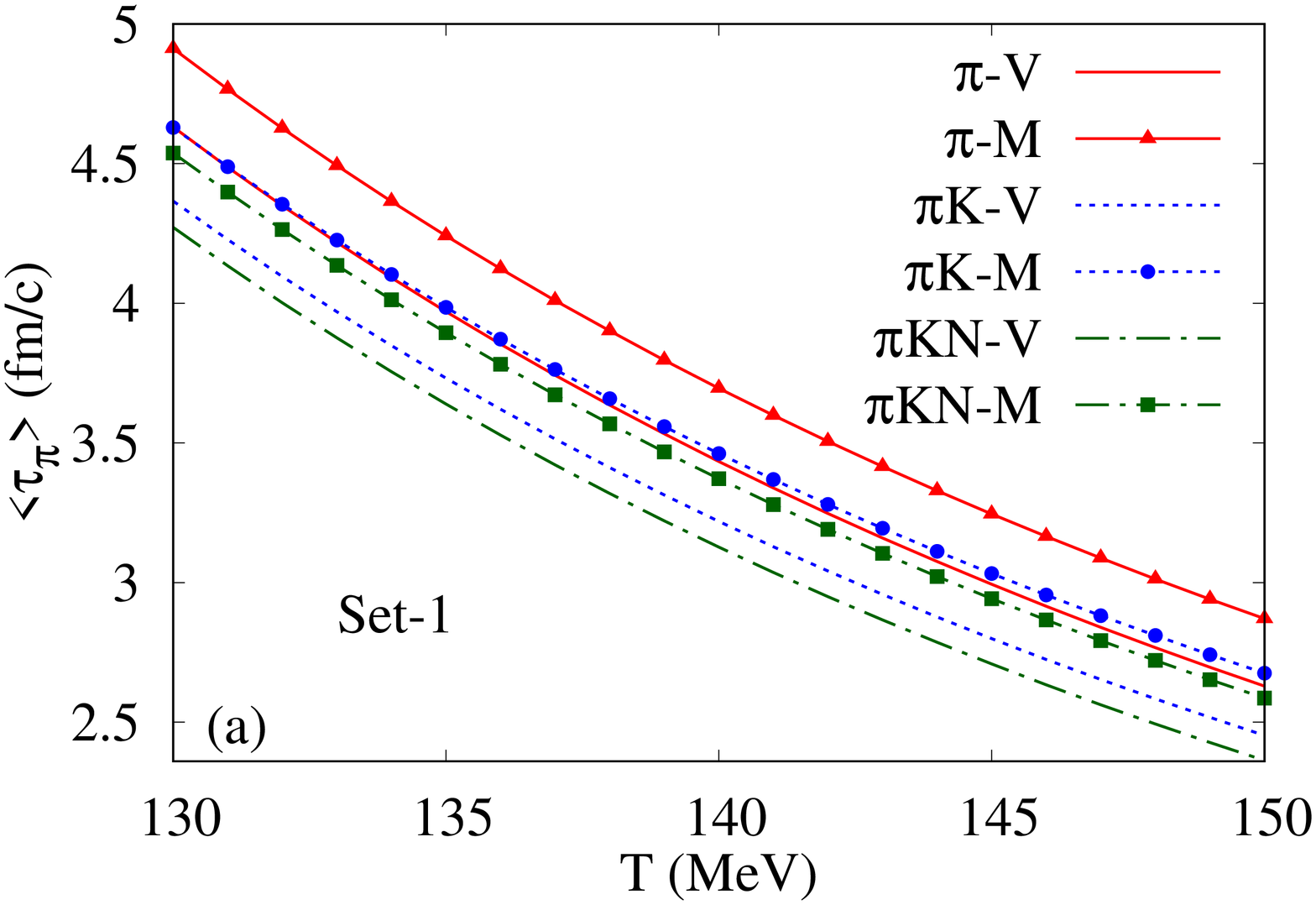}  
		\includegraphics[angle=-90, scale=0.35]{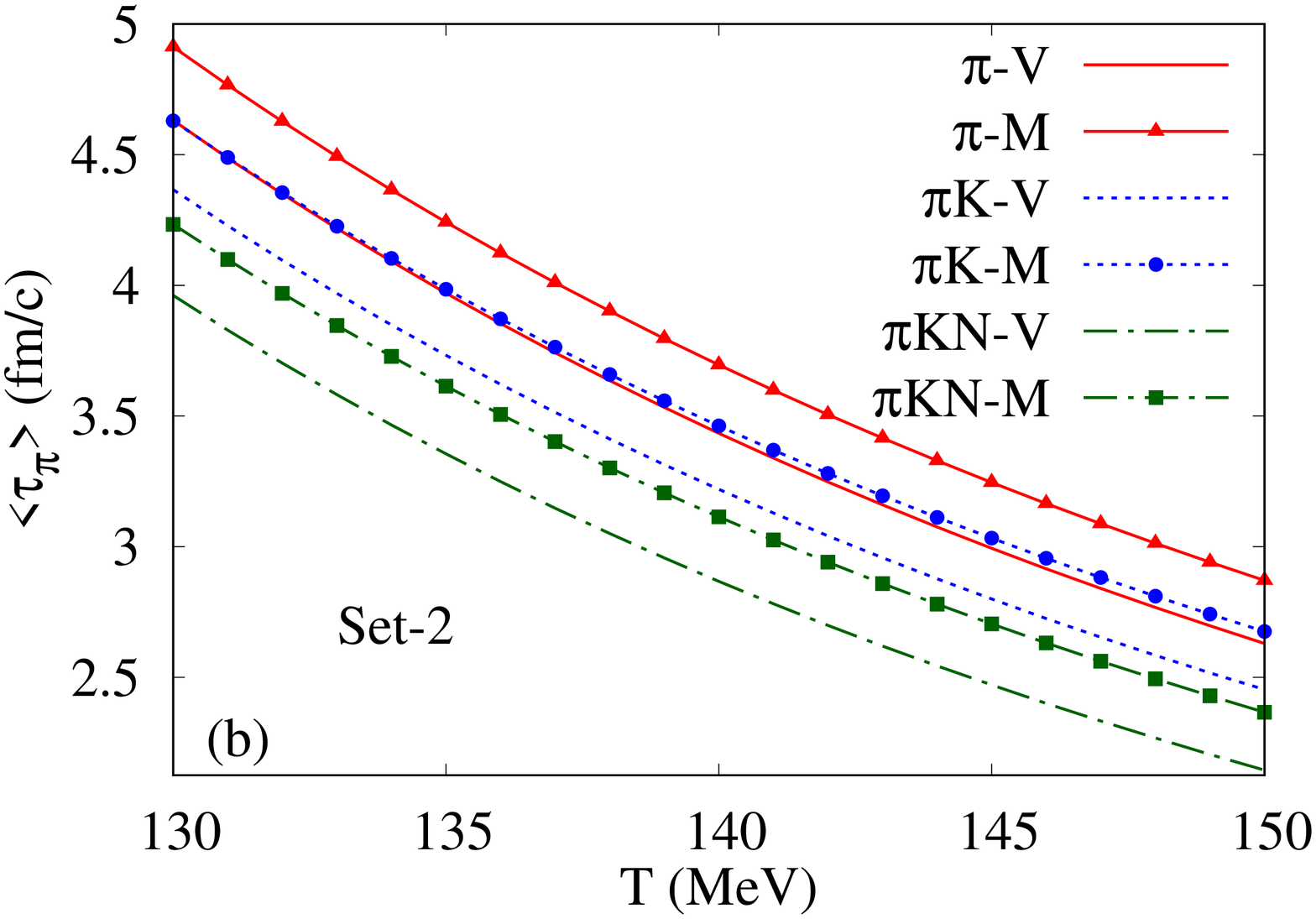} 	
		\includegraphics[angle=-90, scale=0.35]{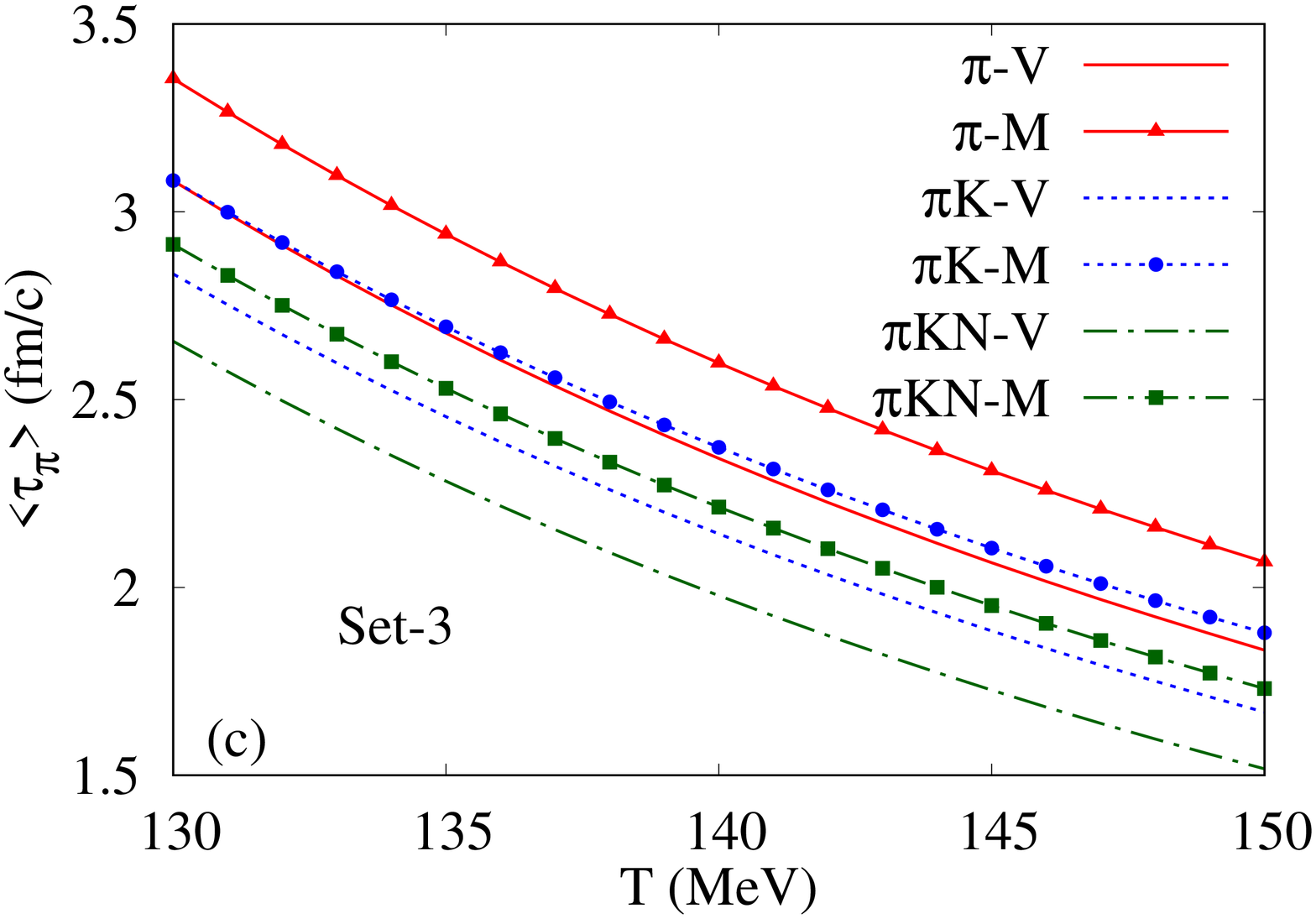}
	\end{center}
	\caption{Mean relaxation time of pions for three different systems under consideration where $V$ and $M$ indicate the use of vacuum and in-medium cross-sections for (a) Set-1, (b) Set-2 and (c) Set-3. }
	\label{tau_avg_pi_diff_comp}
\end{figure}

Here in the upcoming part of this section, we have calculated all the results for three different set of values of pion, nucleon and kaon chemical potential, the choice of these sets have been tabulated in Table~\ref{table1}. In our system the only baryon present is the nucleon and hence baryonic chemical potential is essentially nucleon chemical potential. It is to be noted that non-zero values of $\mu_\pi$ and $\mu_K$ are a consequence of pion and kaon number conservation after chemical freezeout~\cite{Bebie} and the values taken here are demonstrative (see e.g.~\cite{Dobado3}).

Now we turn our attention to the numerical results for the temperature dependence of the momentum averaged relaxation time or collision time of $\pi$, $K$ and $N$. The momentum averaged relaxation time is given by the expression, 
\begin{eqnarray}
\ensembleaverage{\tau_l} = \int d^3p \tau_l f_l \Bigg/ \int d^3p f_l
\end{eqnarray}
where $l\in\{\pi,K,N\}$ and $f_l$ is the equilibrium thermal distribution function of the $l^\text{th}$ species. In all the figures, $V$ and $M$ respectively indicates the use of vacuum and in-medium cross-sections in the calculation of relaxation times.

In Fig.~\ref{tau_avg_pi_diff_comp}, we have shown the average relaxation time of pions as a function of temperature in pion, pion-kaon and pion-kaon-nucleon system with and without medium effects taken into consideration. Essentially there are three noticeable features in the figure. Firstly the decreasing trend of the relaxation time with increasing temperature which can be understood in the following manner. The relaxation time goes like $\sim 1/n\sigma$ where $n$ is the number density and $\sigma$ is the cross section. With the increase in temperature, $n$ increases resulting in a reduction of the relaxation time. Secondly, for a given temperature the system relaxes faster when the number of components rises, as the addition of species increases the net density of particles effectively reducing the mean free path. 
And since the mean free path is directly proportional to the relaxation time, the relaxation time goes down. Finally, we note that the in-medium relaxation times are considerably larger ($ \sim 10-15\% $) compared to their vacuum counterparts. This is due to decrease in cross-section because of the additional scattering and decay processes at finite temperature. The vacuum results are in good agreement with Ref.~\cite{Prakash}. From the different sets we see that with the increase in chemical potential the magnitude of average relaxation time of pions have decreased.

\begin{figure}[h]
	\begin{center}
		\includegraphics[angle=-90, scale=0.35]{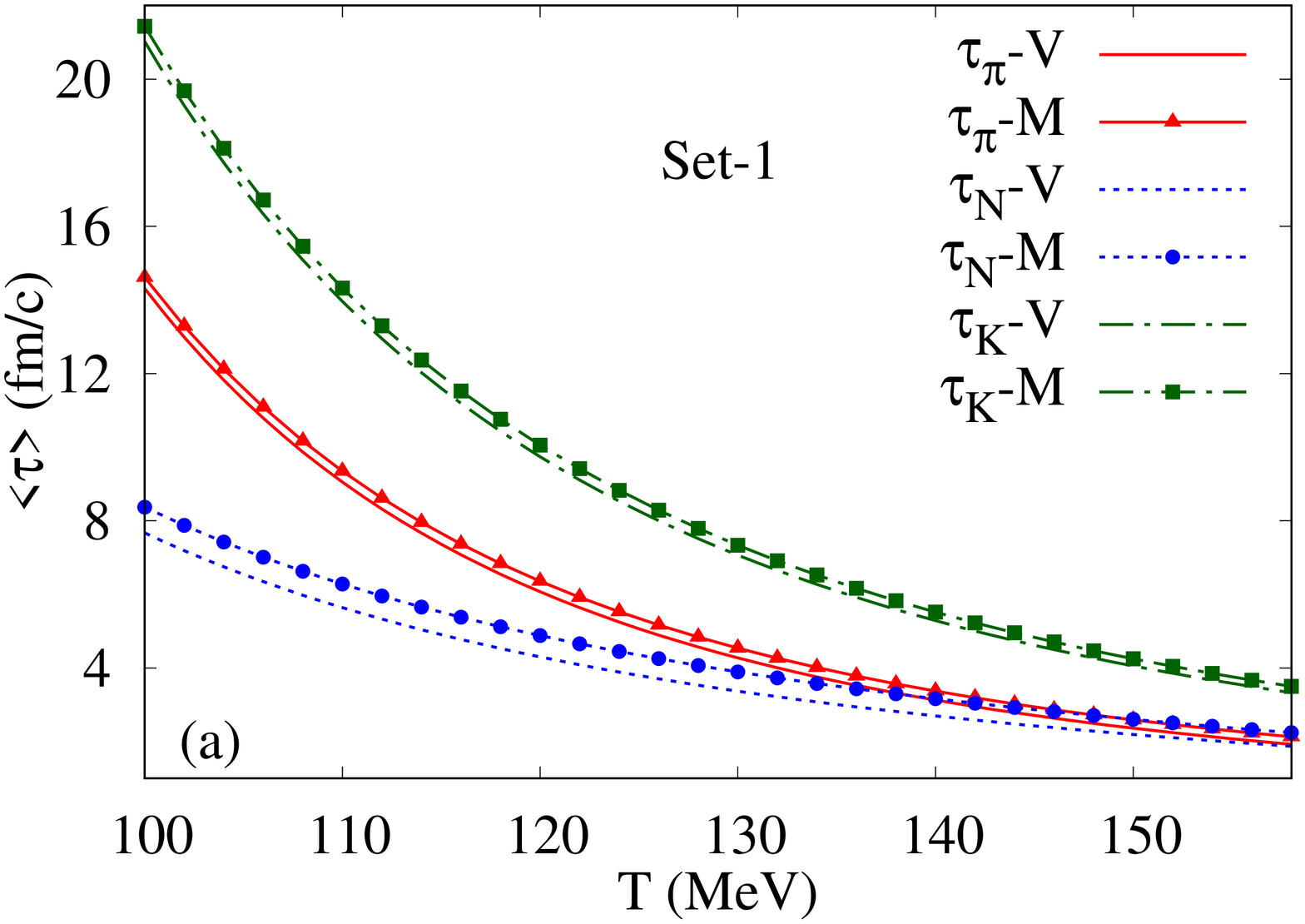}  
		\includegraphics[angle=-90, scale=0.35]{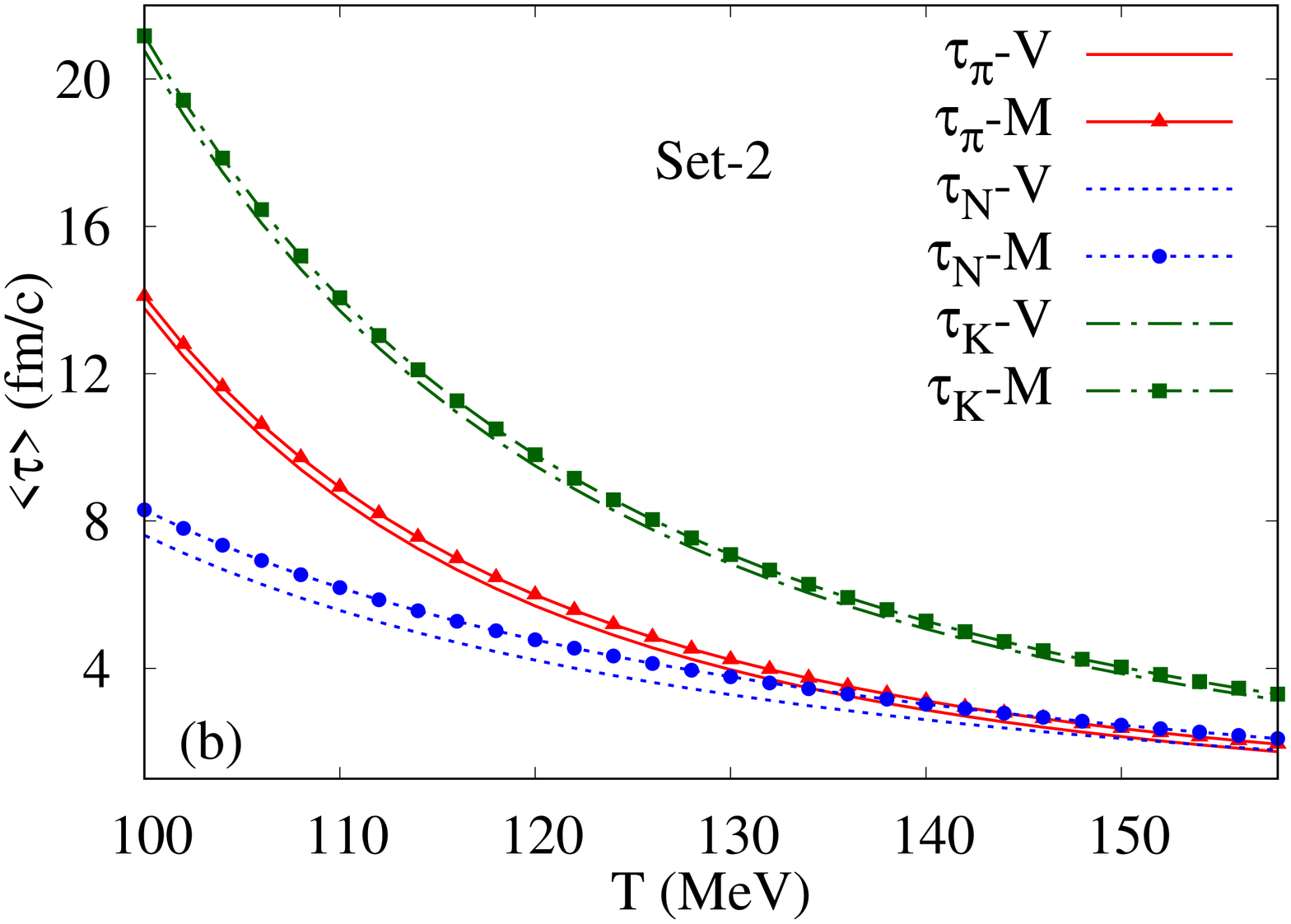} 
		\includegraphics[angle=-90, scale=0.35]{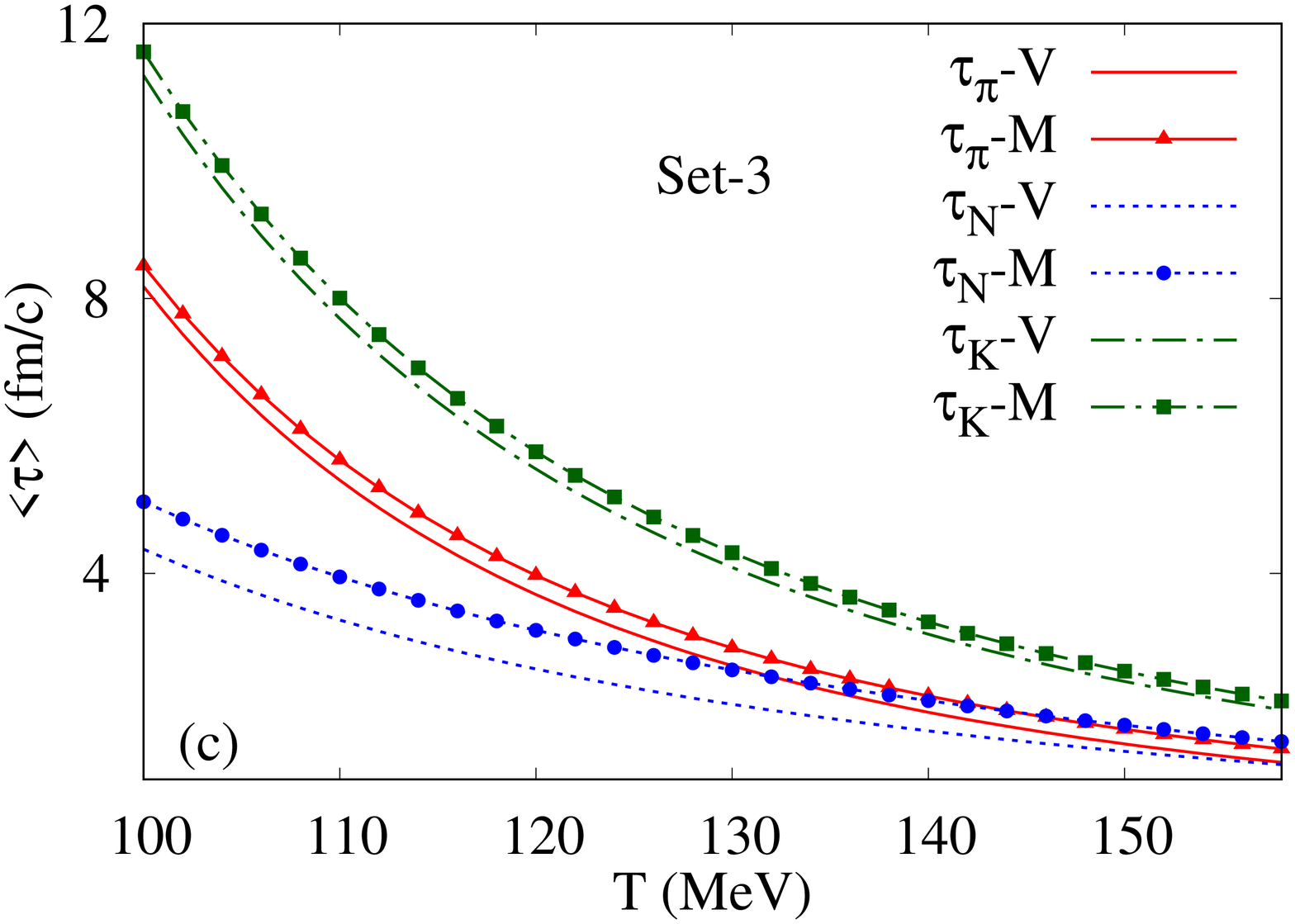}
	\end{center}
	\caption{Momentum averaged relaxation time of pions, nucleons and kaons in a pion-nucleon-kaon hadronic gas as function of temperature for (a) Set-1, (b) Set-2 and (c) Set-3 of chemical potentials of individual components.
	}
	\label{Fig_tau_vs_T}
\end{figure}

In Figs.~\ref{Fig_tau_vs_T}(a)-(c) 
we have presented the dependence of mean relaxation time of individual species of the hadron gas mixture, i.e. pion, kaons and nucleons, on temperature 
with and without medium effects. It is noted that the mean relaxation time of kaon remains larger compared to the other constituents of the system in all cases over the temperature range considered because of their smaller cross-section. From Eq.~(\ref{relax}), it is evident that mean relaxation time of the components are interdependent. With increasing baryonic density, the relaxation mechanism in the system gets enhanced as the number of particles with which collision is possible increases, and hence a relative decrease in the relaxation time as seen in Figs.~\ref{Fig_tau_vs_T}(a)-(c). The in-medium behaviour of the relaxation time can also be explained along similar lines as done before.

\begin{figure}[h]
	\begin{center}
		\includegraphics[angle=-90, scale=0.35]{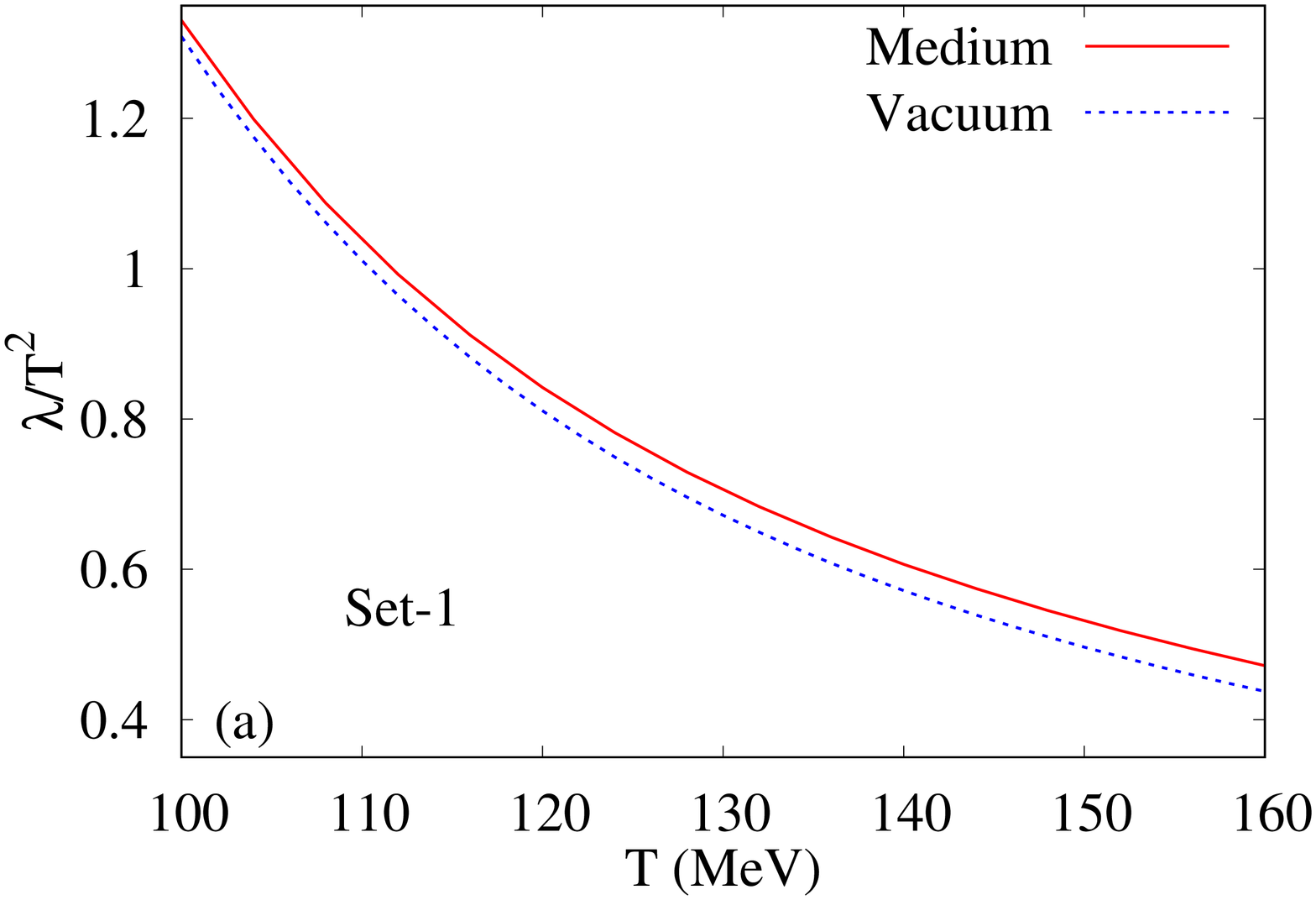}  
		\includegraphics[angle=-90, scale=0.35]{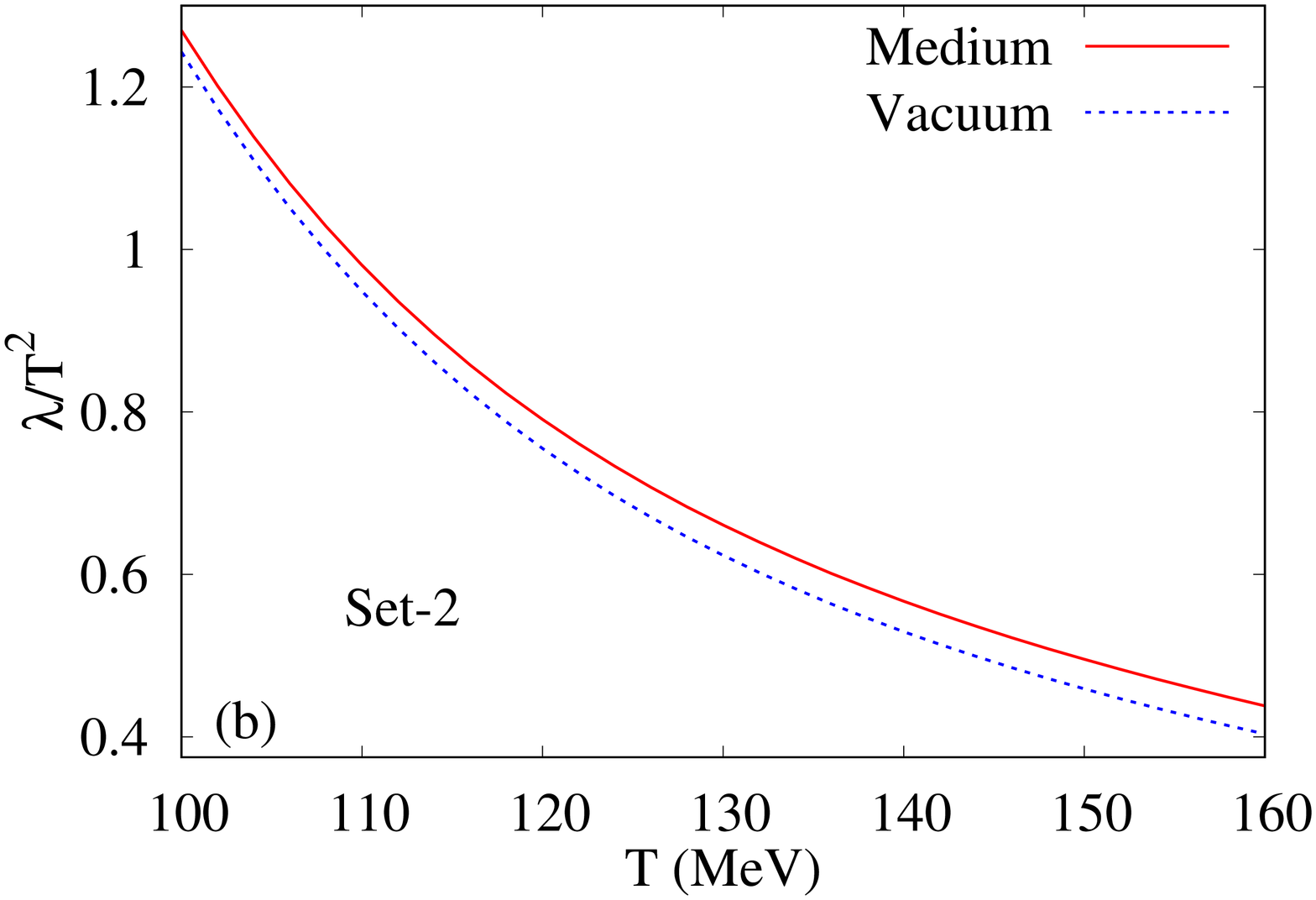} 	
		\includegraphics[angle=-90, scale=0.35]{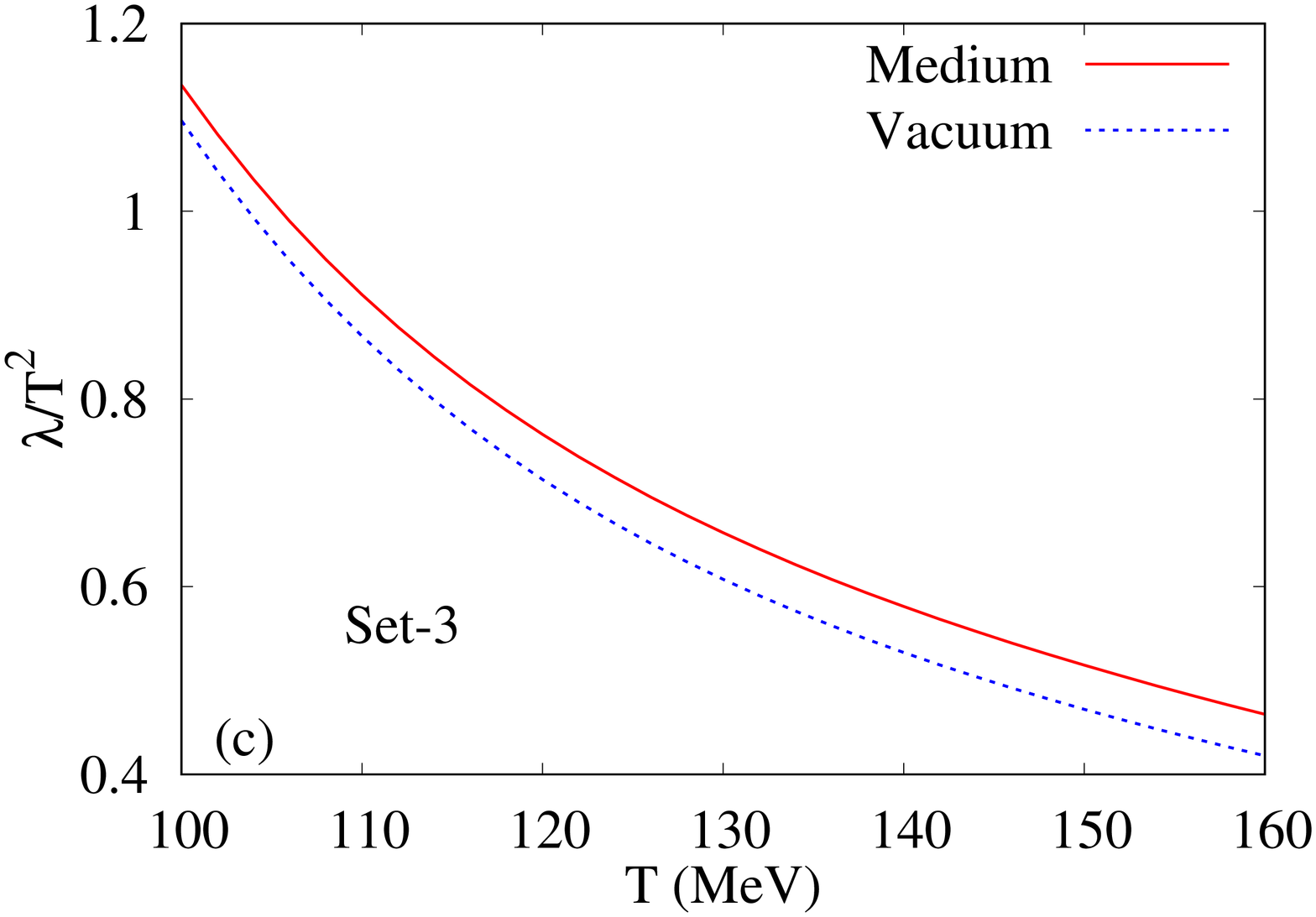}
	\end{center}
	\caption{$\lambda/T^{2}$ as a function of temperature for (a) Set-1, (b) Set-2 and (c) Set-3 of chemical potentials of individual components. }
	\label{Fig_lambda_vs_T}
\end{figure}
\begin{figure}[h]
	\begin{center}
		\includegraphics[angle=-90, scale=0.35]{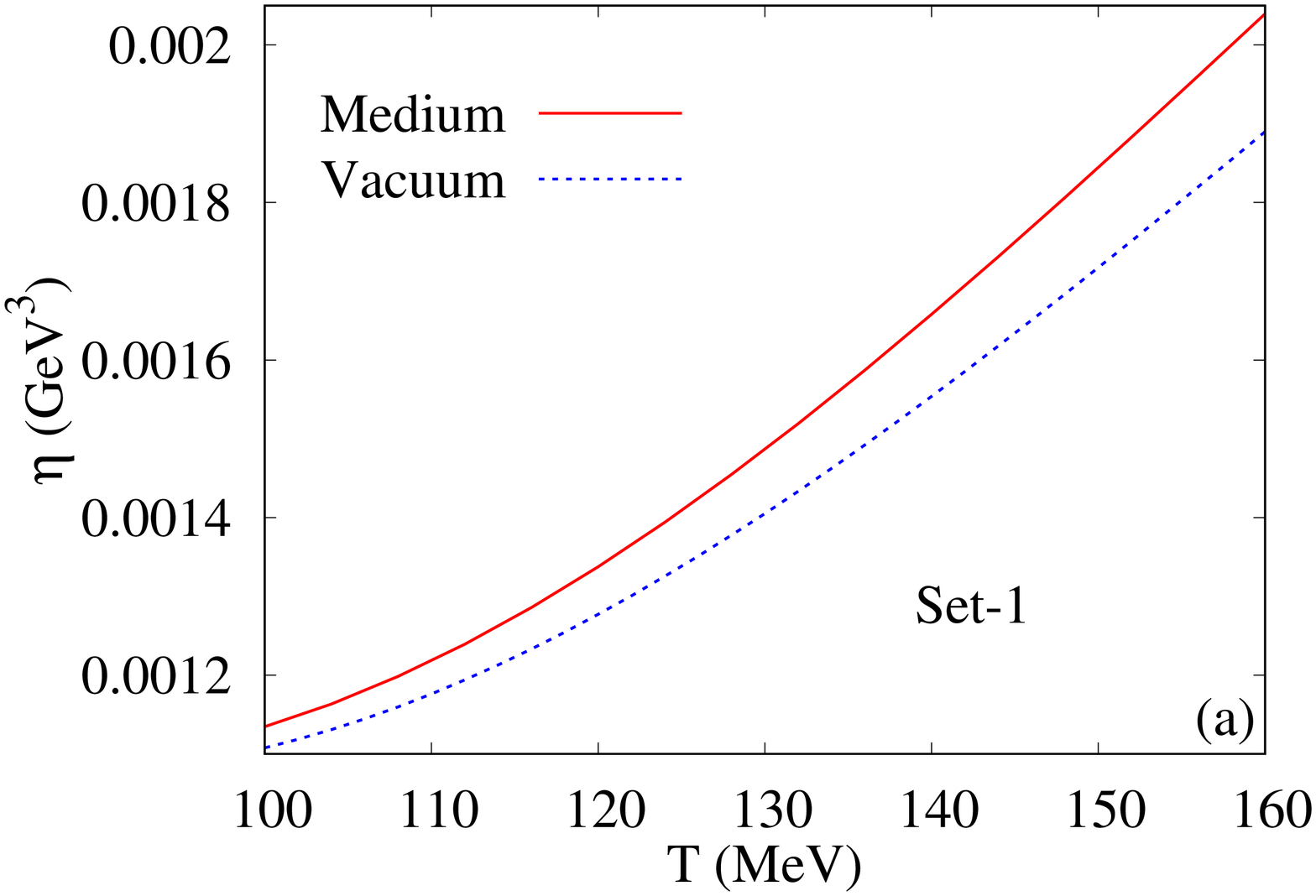}  
		\includegraphics[angle=-90, scale=0.35]{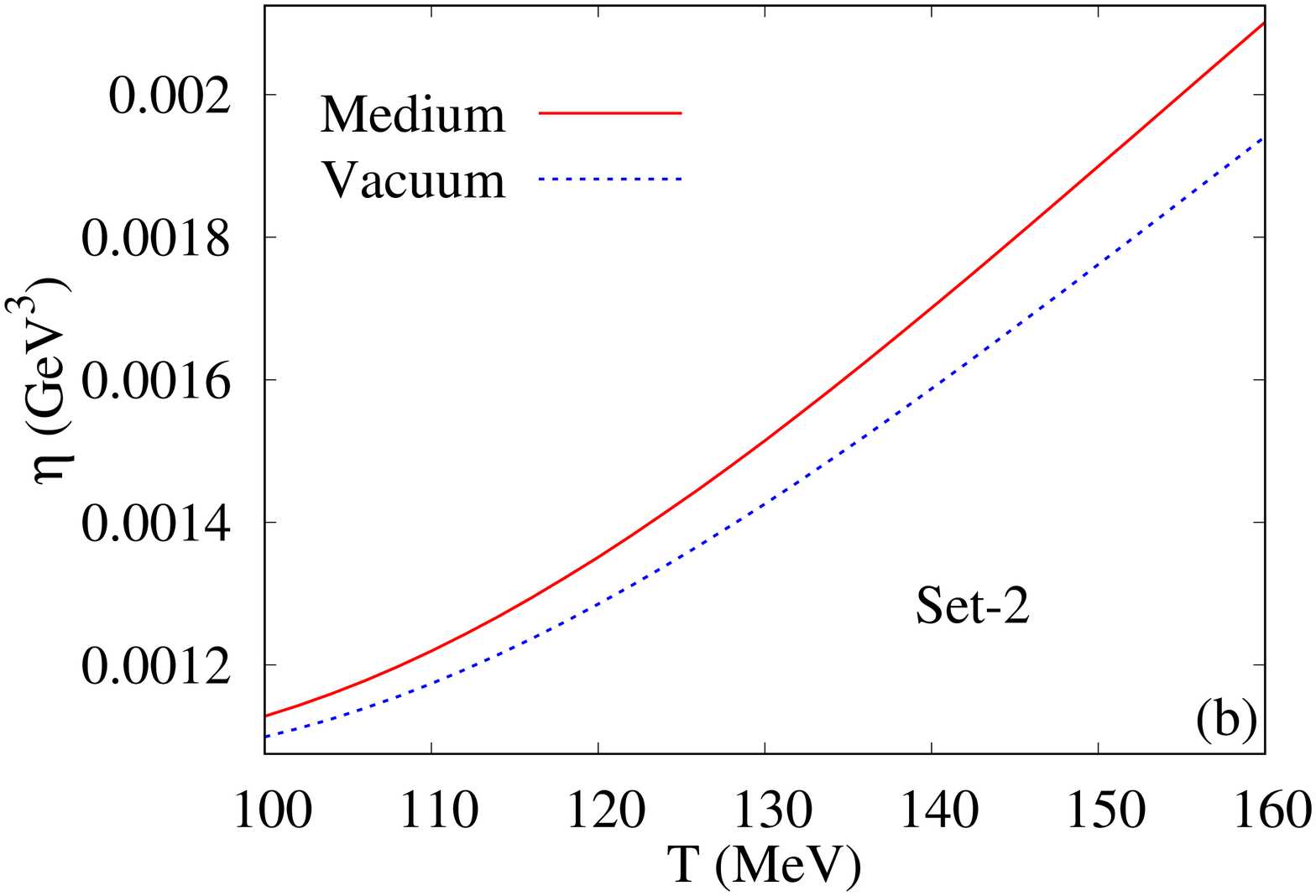} 
		\includegraphics[angle=-90, scale=0.35]{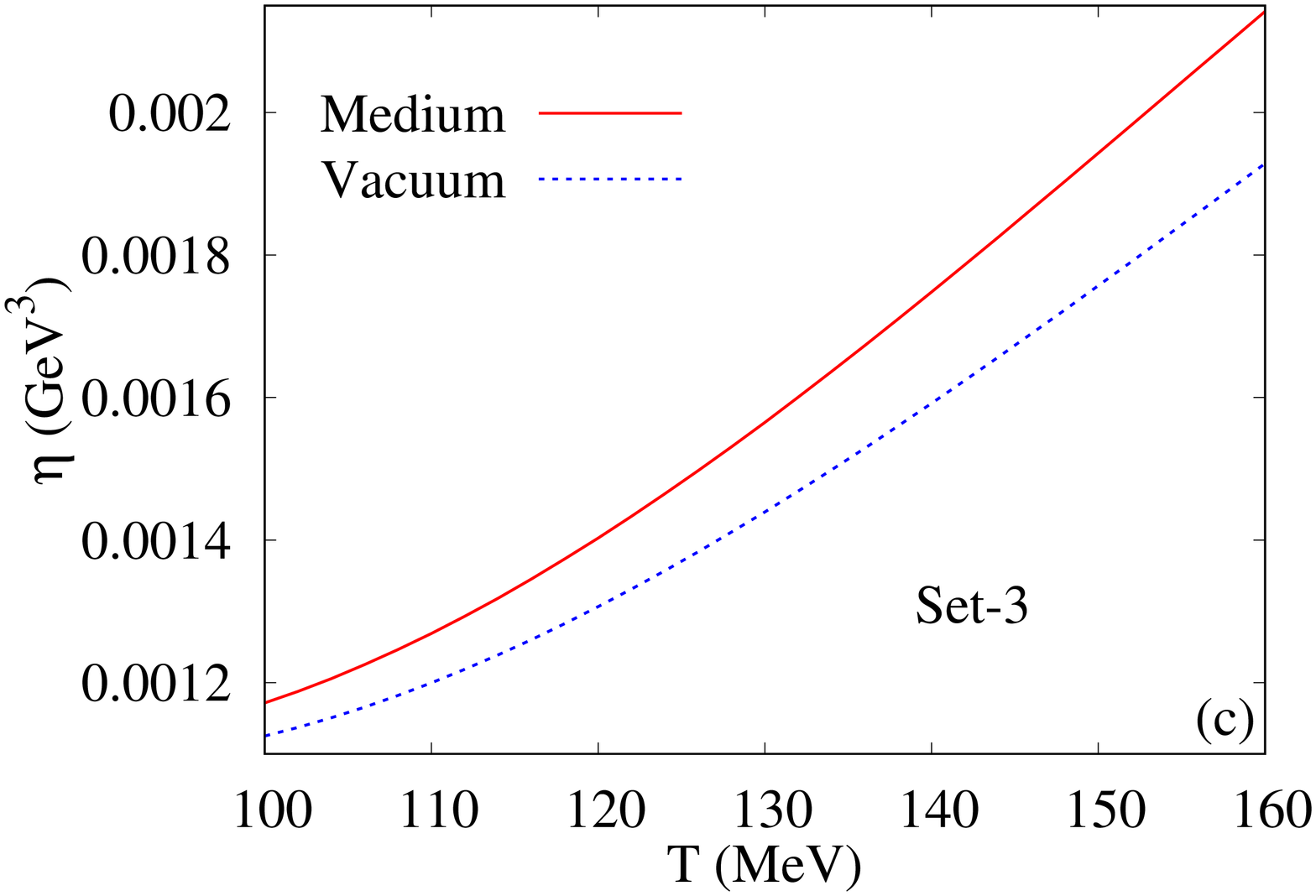} 
		\includegraphics[angle=-90, scale=0.35]{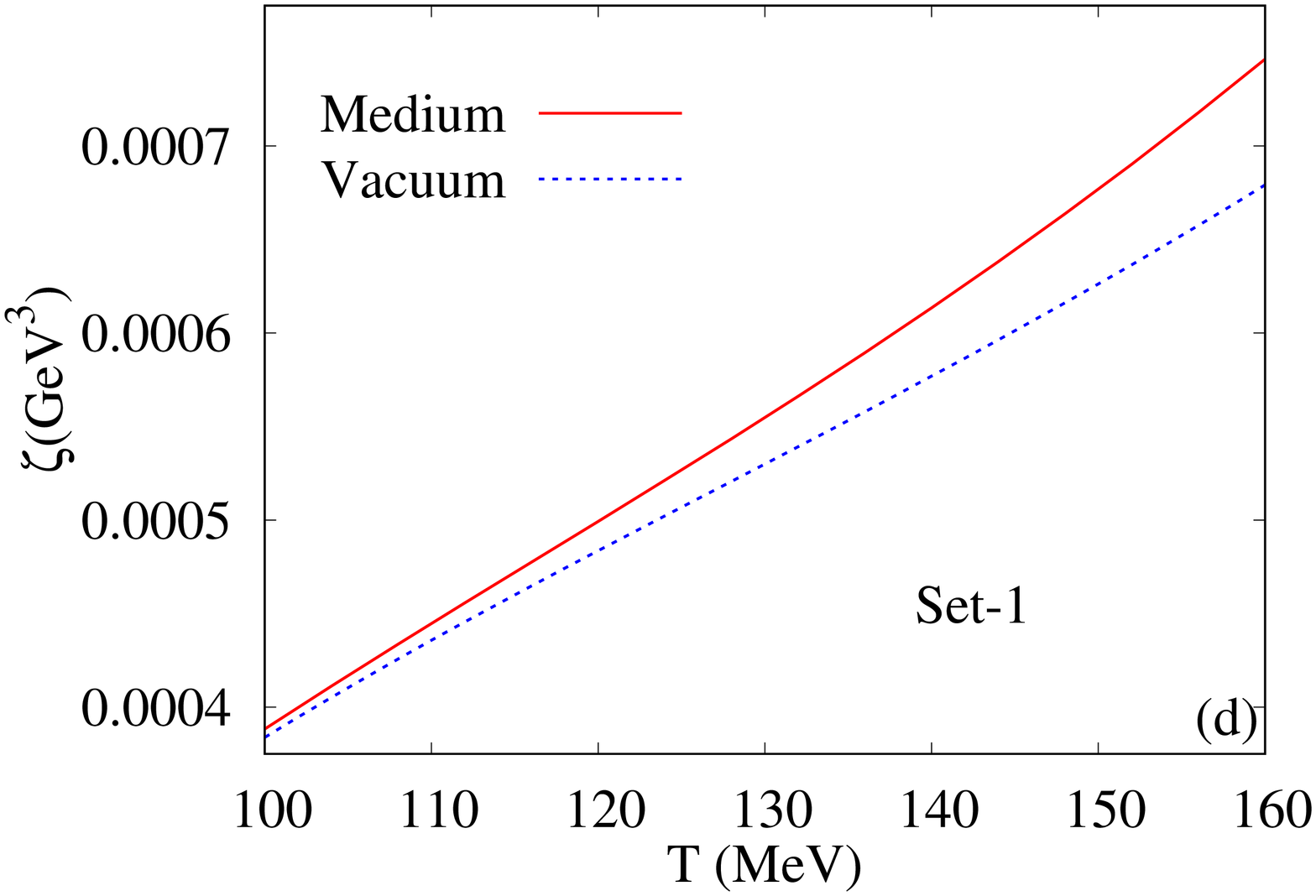}  
		\includegraphics[angle=-90, scale=0.35]{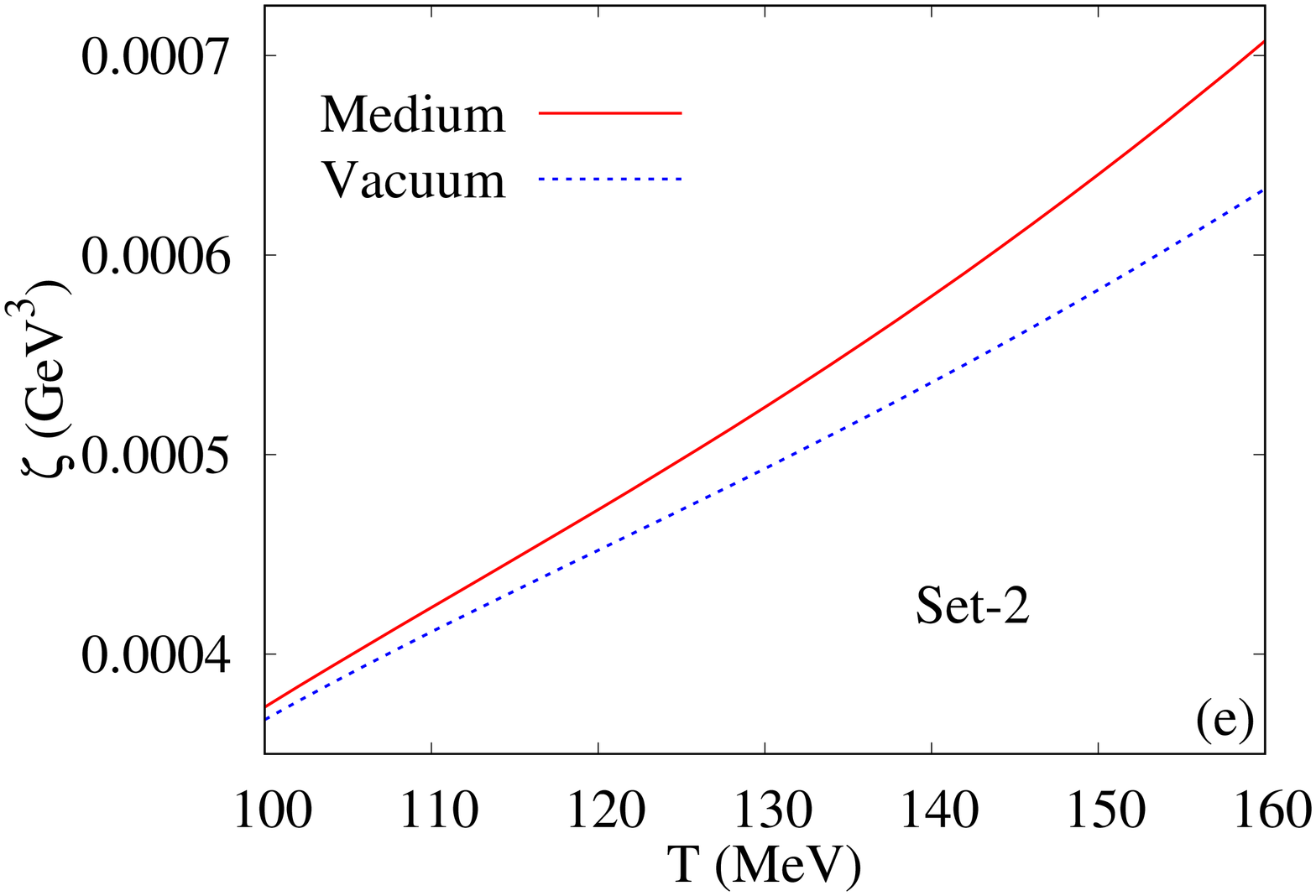} 	
		\includegraphics[angle=-90, scale=0.35]{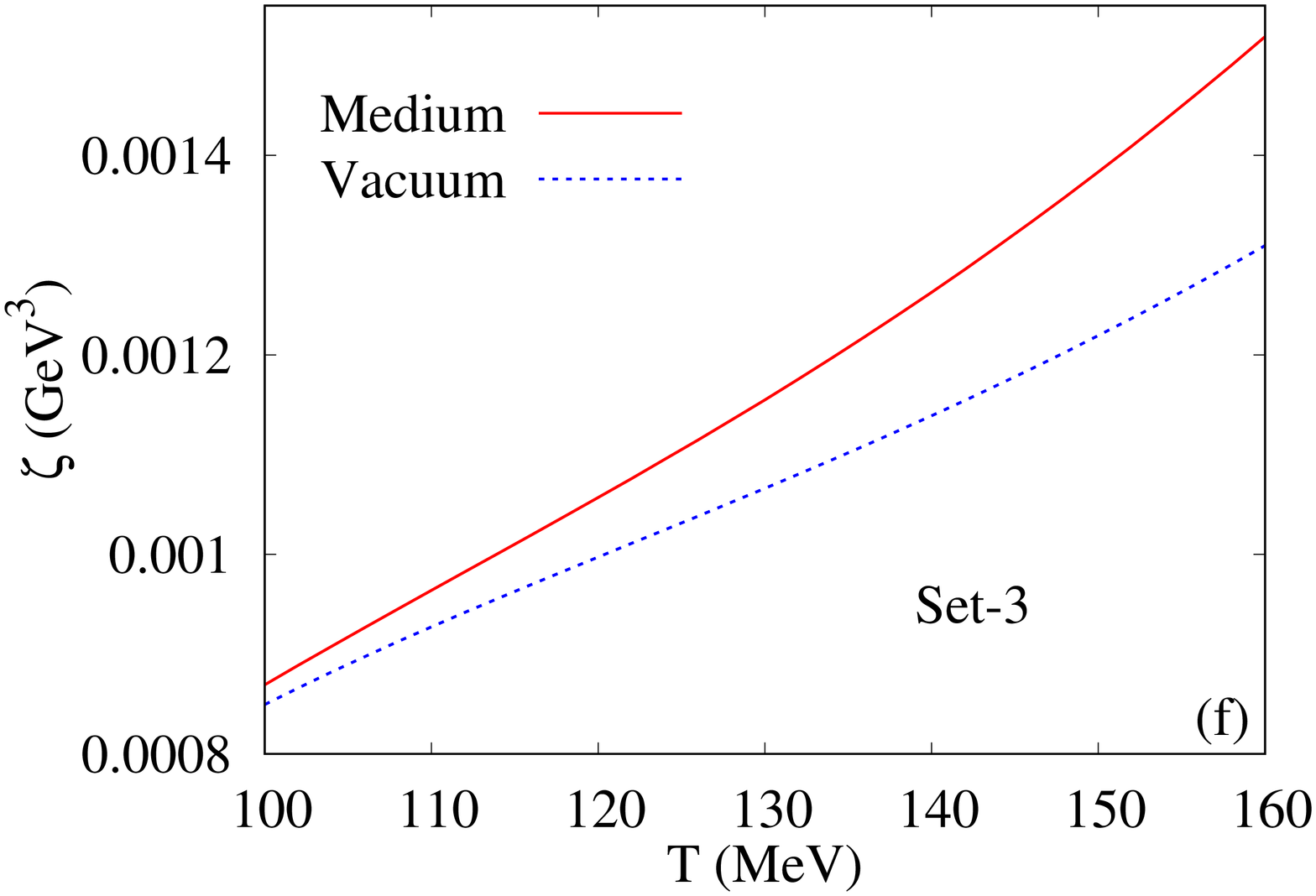}
	\end{center}
	\caption{Shear viscosity($ \eta $) and bulk viscosity ($\zeta$) as a function of temperature ($T$) for a pion-kaon-nucleon hadronic gas mixture for (a) Set-1, (b) Set-2 and (c) Set-3 of chemical potentials of individual components with and without including medium effects.}
	\label{Fig_etazeta_vs_T}
\end{figure}

Having been studied the behaviour of the relaxation times of different species, we now turn our attention to the transport coefficients. 
The temperature dependence of scaled thermal conductivity $\lambda/T^2$ for different sets of chemical potential of the constituents is shown in Figs.~\ref{Fig_lambda_vs_T}(a)-(c). The quantity $\lambda/T^2$ decreases with increase in temperature and also decreases with the increase in chemical potential of the constituents. The fall of $\lambda/T^2$ is similar to that of relaxation time, steeper at lower temperature and gradual as temperature increases. Medium effects are reflected by the increase in magnitude of $\lambda/{T^2}$. With the increase in chemical potential the density of the heavier particles like nucleons and kaons increases which brings down the relaxation time, thus reducing the thermal conductivity.

We now proceed to present the numerical evaluation of shear and bulk viscosity using Eqs.~(\ref{eta}) and (\ref{zeta}). The results are shown for the temperature range starting from $100$ to $160$ MeV which is typical of a hadron gas produced in the later stages of heavy ion collisions.  In Figs.~\ref{Fig_etazeta_vs_T}(a)-(c) and Figs.~\ref{Fig_etazeta_vs_T}(d)-(f), the vacuum and in-medium evolution of shear viscosity ($\eta$) and bulk viscosity ($\zeta$) as a function of temperature are shown. Classically, for a single component gas, one can write $ \eta \propto \bar{p}/\sigma $, where $ \bar{p} $ and $ \sigma $ are the average momentum and binary cross section respectively. Fig.~\ref{cross_sec} shows that the cross-section reduces with increasing temperature ($T$) and since $\bar{p}$ goes as $\sqrt{T}$, the rise in magnitude of $\eta$  due to inclusion of medium effects is understandable from the expression of $\eta$.  
The medium effects are reflected in the increase in magnitude of $\zeta$ as compared to that of $\zeta$ calculated using vacuum cross-sections.

Let us now proceed to study the behaviour of the ratios of viscosities to the entropy density $\eta/s$ and $\zeta/s$ which are also termed as the specific shear and bulk viscosity. For this, we need the expression of the entropy density. First we note that, the entropy density of a  non-interacting hadronic gas mixture is given by~\cite{Venugopalan:1992hy}
\begin{eqnarray}
s_\text{free} = T^3\sum_{h\in\{\text{hadrons}\}} \frac{g_h}{2\pi^2}  \FB{\frac{m_h}{T}}^2\TB{\FB{\frac{m_h}{T}}\mathcal{S}_{h3}^1\FB{\frac{m_h}{T},\frac{\mu_h}{T}}
-\FB{\frac{\mu_h}{T}}\mathcal{S}_{h2}^1\FB{\frac{m_h}{T},\frac{\mu_h}{T}}}
\end{eqnarray}
where, the sum runs over all the hadronic species taken into consideration, $m_h$ is the mass, $g_h$ is the spin-isospin degeneracy, 
$\mu_h$ is the chemical potential of the hadron $h$ and 
\begin{eqnarray}
\mathcal{S}_{hn}^\alpha(x,y) = \sum_{j=1}^{\infty}(a_h)^{j+1} e^{jy} j^{-\alpha} K_n(jx)
\end{eqnarray}
in which $a_h=1$ if $h$ is a Boson and $a_h=-1$ if $h$ is a Fermion. 
\begin{figure}[h]
	\begin{center}
		\includegraphics[angle=-90, scale=0.35]{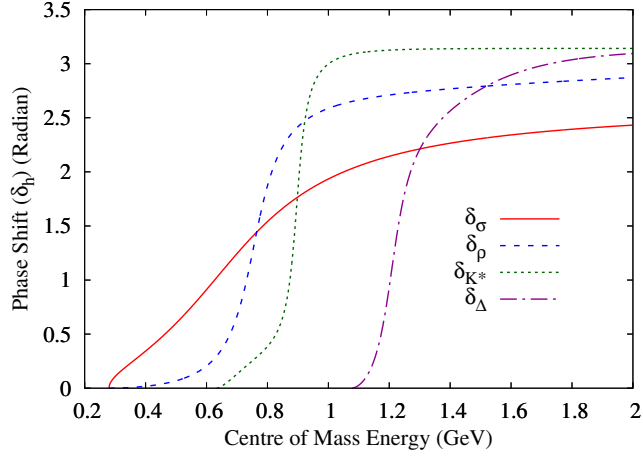}  
	\end{center}
	\caption{Phase shifts in different resonance channels as a function of center of mass energy. }
	\label{fig:phase}
\end{figure}
\begin{figure}[h]
	\begin{center}
		\includegraphics[angle=-90, scale=0.35]{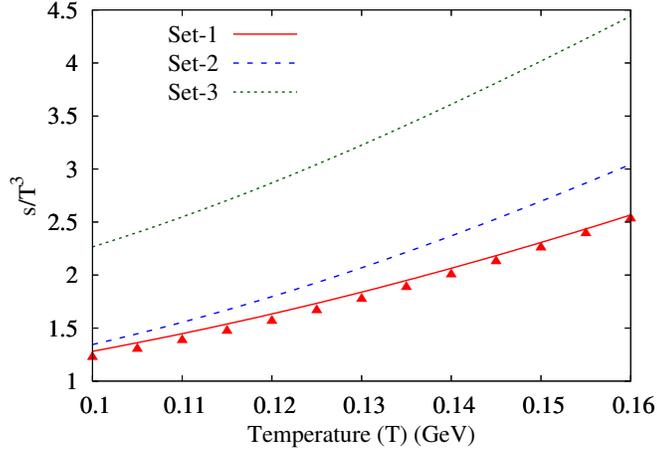}  
	\end{center}
	\caption{Entropy density scaled by the cube of temperature ($s/T^3$) as a function of temperature for different sets of chemical potentials. The triangles correspond to the result for a free gas mixture of $\pi, K, N, \sigma, \rho, K^*$ and $\Delta$.}
	\label{fig:entropy}
\end{figure}

In order to take into account the effect of interactions among the different hadrons, we use the relativistic virial expansion as discussed in Refs.~\cite{Venugopalan:1992hy,Wiranata:2013oaa}. The total entropy then comes out to be the sum of the free and interacting parts $s = s_\text{free} + s_\text{int}$. 
The leading contribution to $s_\text{int}$ comes from the second virial coefficient which can be calculated from two body phase shifts as
\begin{eqnarray}
s_\text{int} = \sum_{i,j\ge i}e^{\mu_i/T}e^{\mu_j/T} \frac{1}{2\pi^3}\int_{m_i+m_j}^{\infty}d(\sqrt{s}) s
\TB{\frac{\sqrt{s}}{T} K_2\FB{\frac{\sqrt{s}}{T}}-\FB{\frac{\mu_i}{T}+\frac{\mu_j}{T}}K_2\FB{\frac{\sqrt{s}}{T}}}
\sum_{c\in\text{channels}} g_c\delta_c^{ij}
\end{eqnarray}
where the indices $i,j$ run over all the hadronic species taken into consideration, $\sqrt{s}$ is the center of mass energy for the hadronic pair $\{i,j\}$, $g_c$ is the spin-isospin degeneracy of the resonance being exchanged in the channel $c$ and $\delta_c^{ij}$ is the corresponding phase shift. The dominant contribution to $s_\text{int}$ come from those channels where explicit resonance exchanges occur~\cite{Venugopalan:1992hy}. For consistency with the calculations of in-medium cross-sections here we only take into account the resonance channels considered in $s$-channel scattering matrix elements described in Section III. The phase shift ($\delta_R$) for the resonance ($R$) exchange can be obtained from
\begin{eqnarray}
\delta_R(s) = \frac{\pi}{2} + \tan^{-1}\FB{\frac{\sqrt{s}-m_R}{\Gamma_R(s)/2}}
\end{eqnarray}
where, $\Gamma_R(s)$ is the decay width of the resonance $R$ with four-momentum $(\sqrt{s},\vec{0})$. 
\begin{figure}[h]
	\begin{center}
		\includegraphics[angle=-90, scale=0.35]{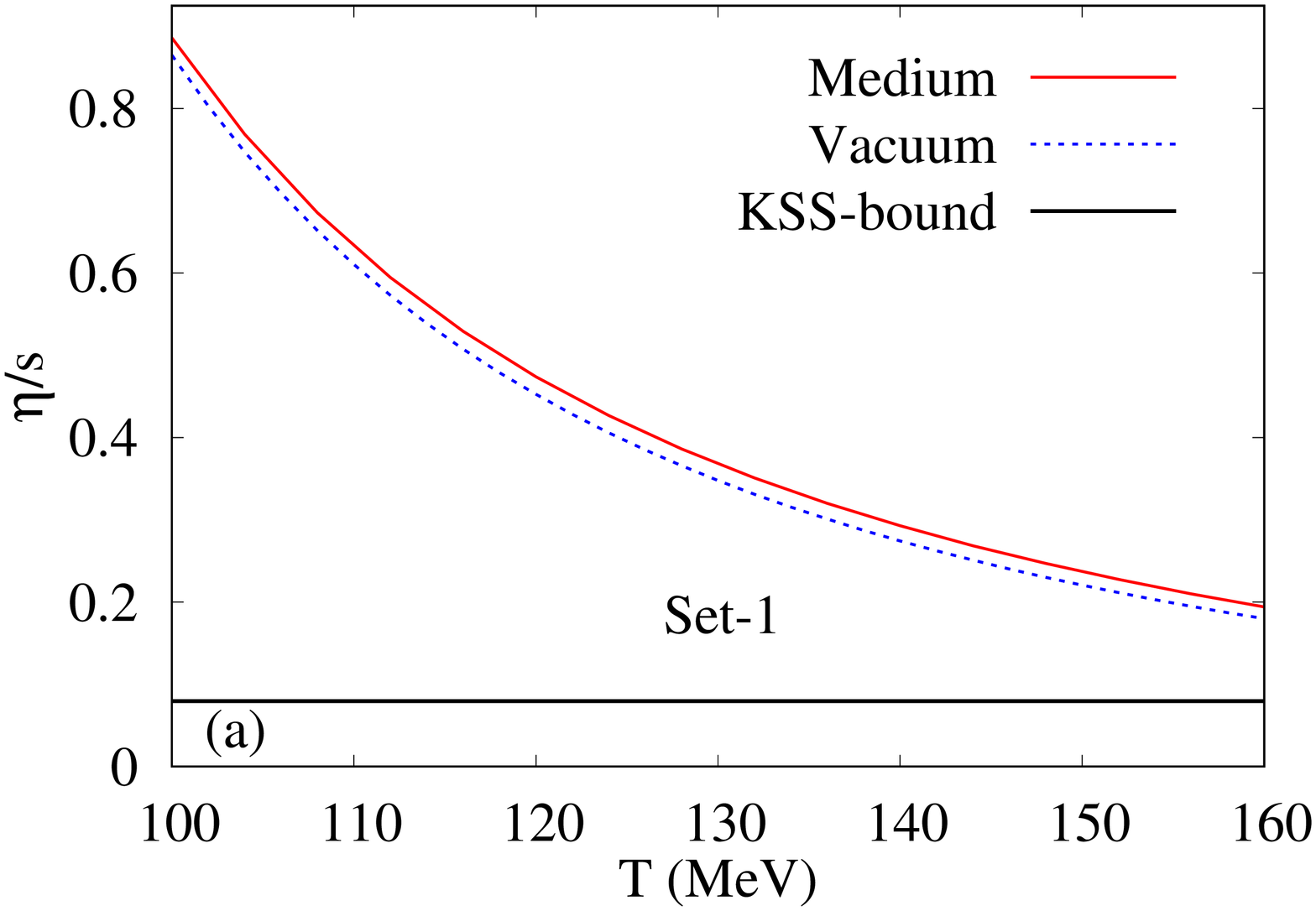}  
		\includegraphics[angle=-90, scale=0.35]{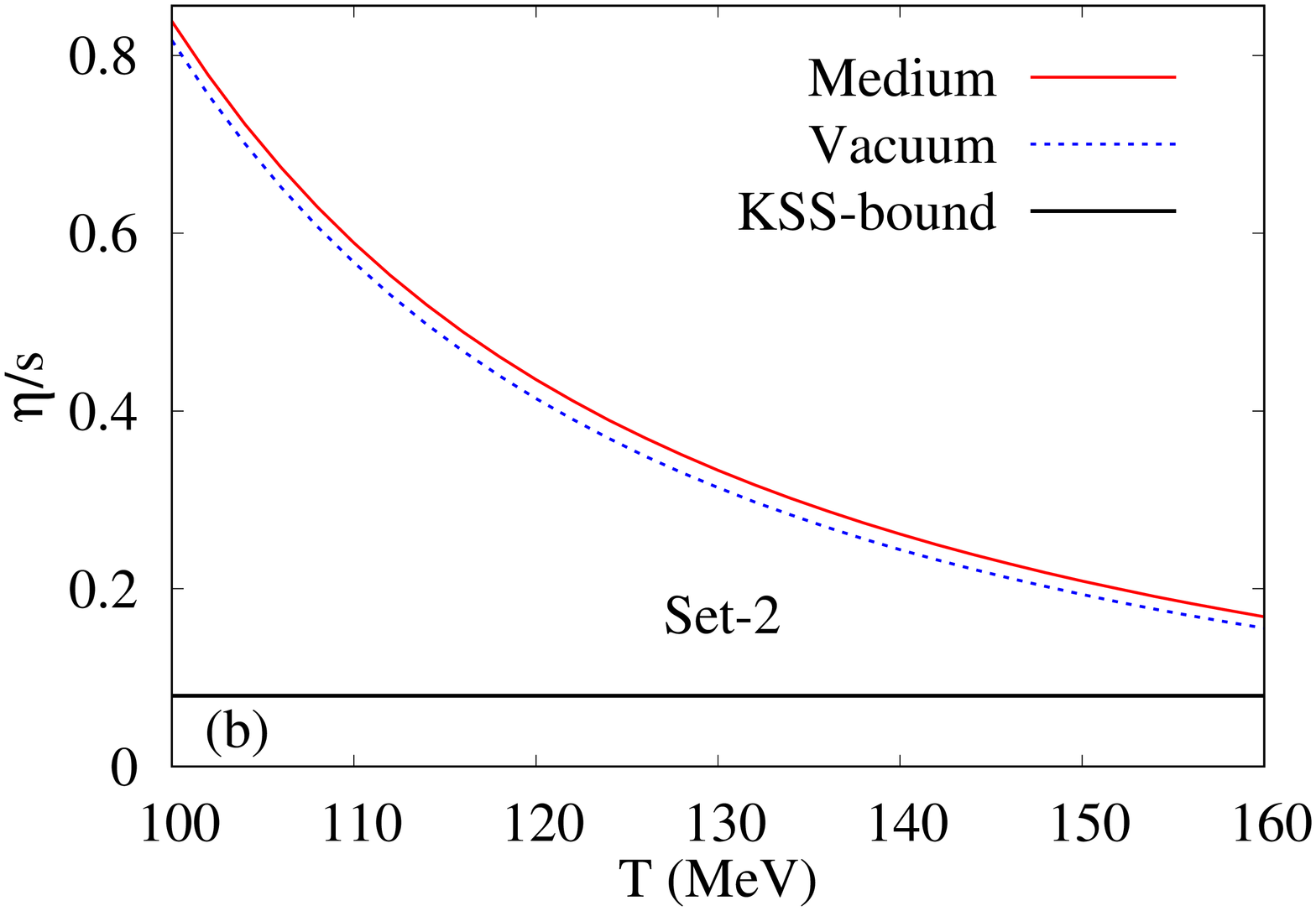} 	
		\includegraphics[angle=-90, scale=0.35]{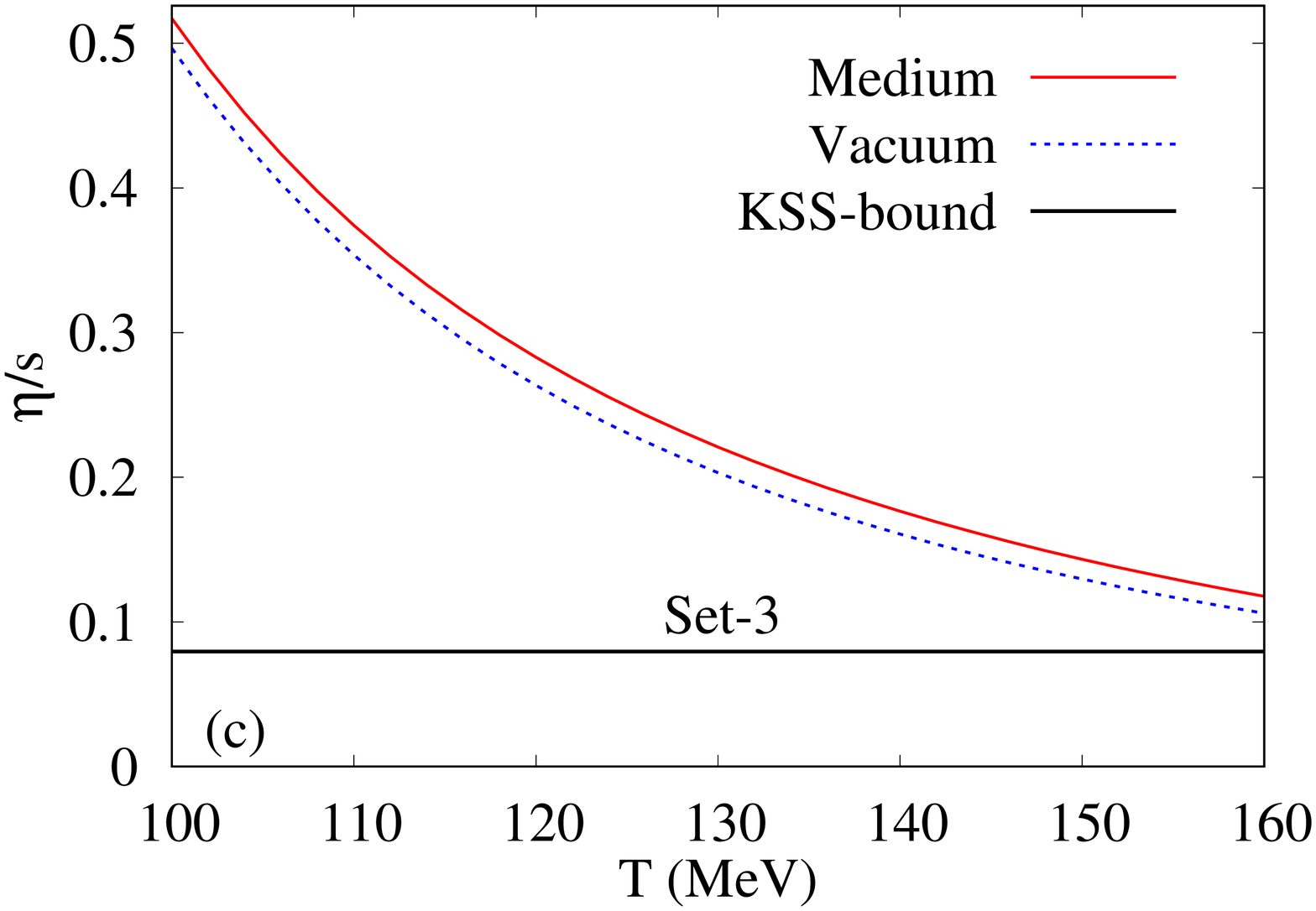}
		\includegraphics[angle=-90, scale=0.35]{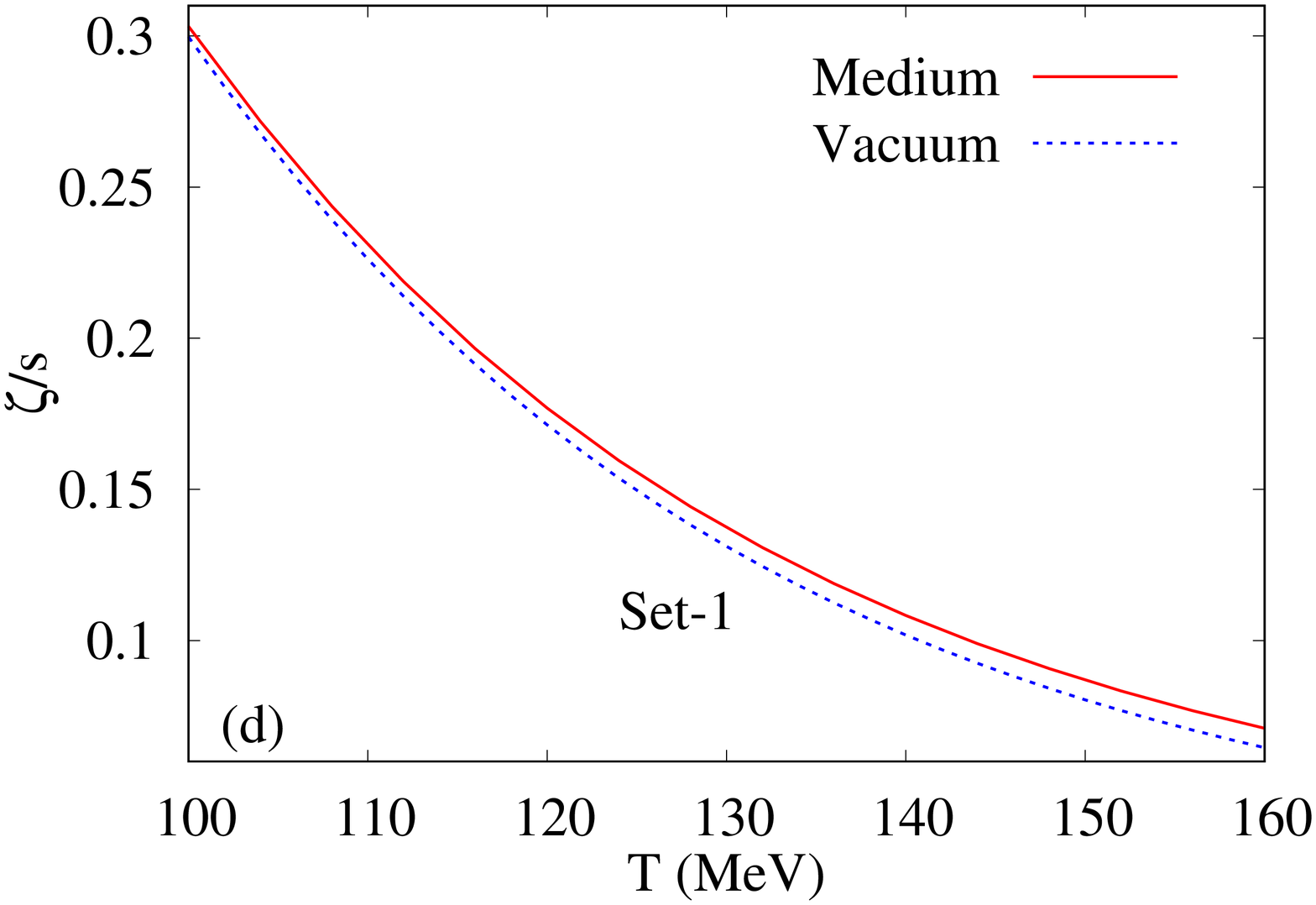}  
		\includegraphics[angle=-90, scale=0.35]{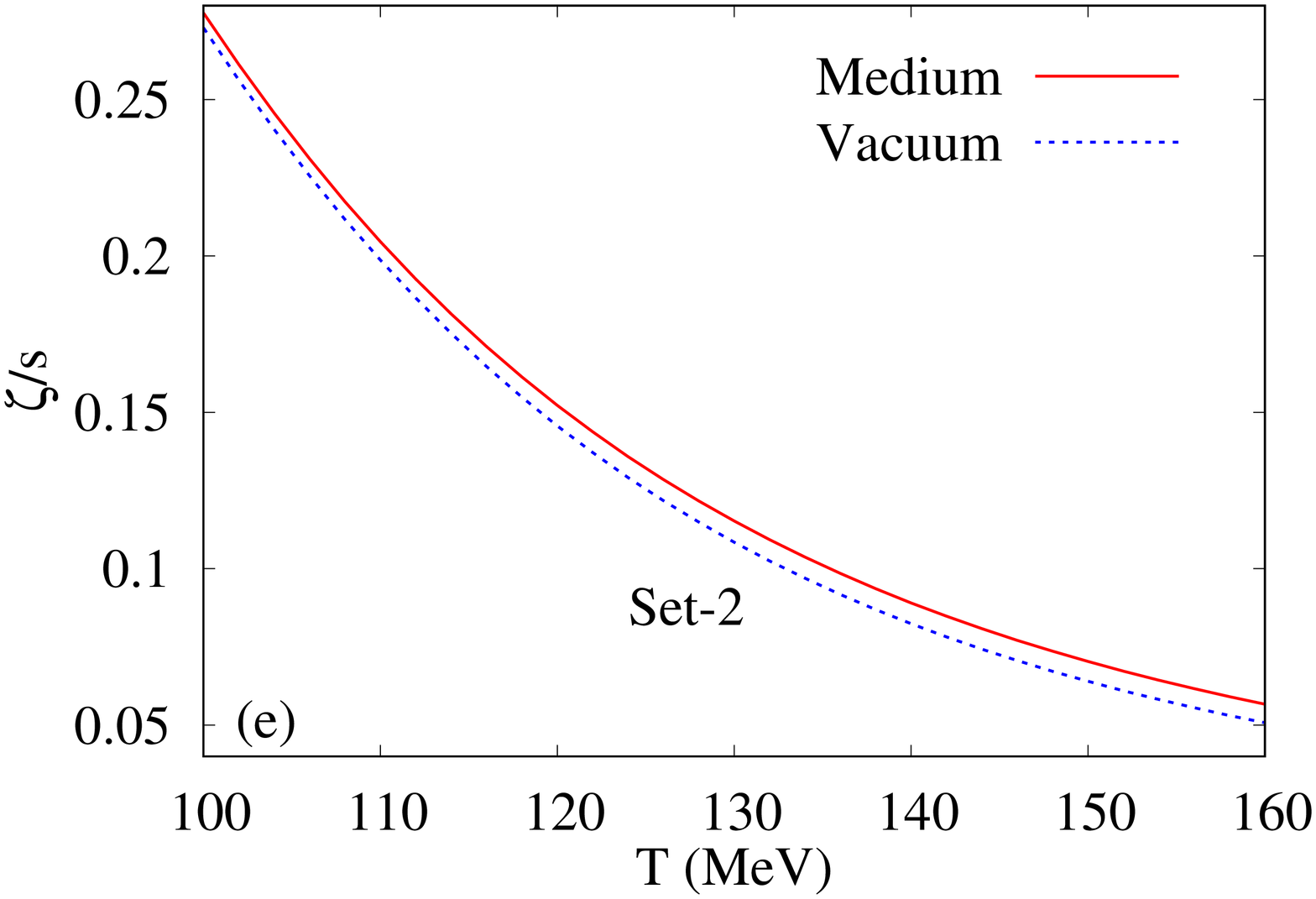} 	
		\includegraphics[angle=-90, scale=0.35]{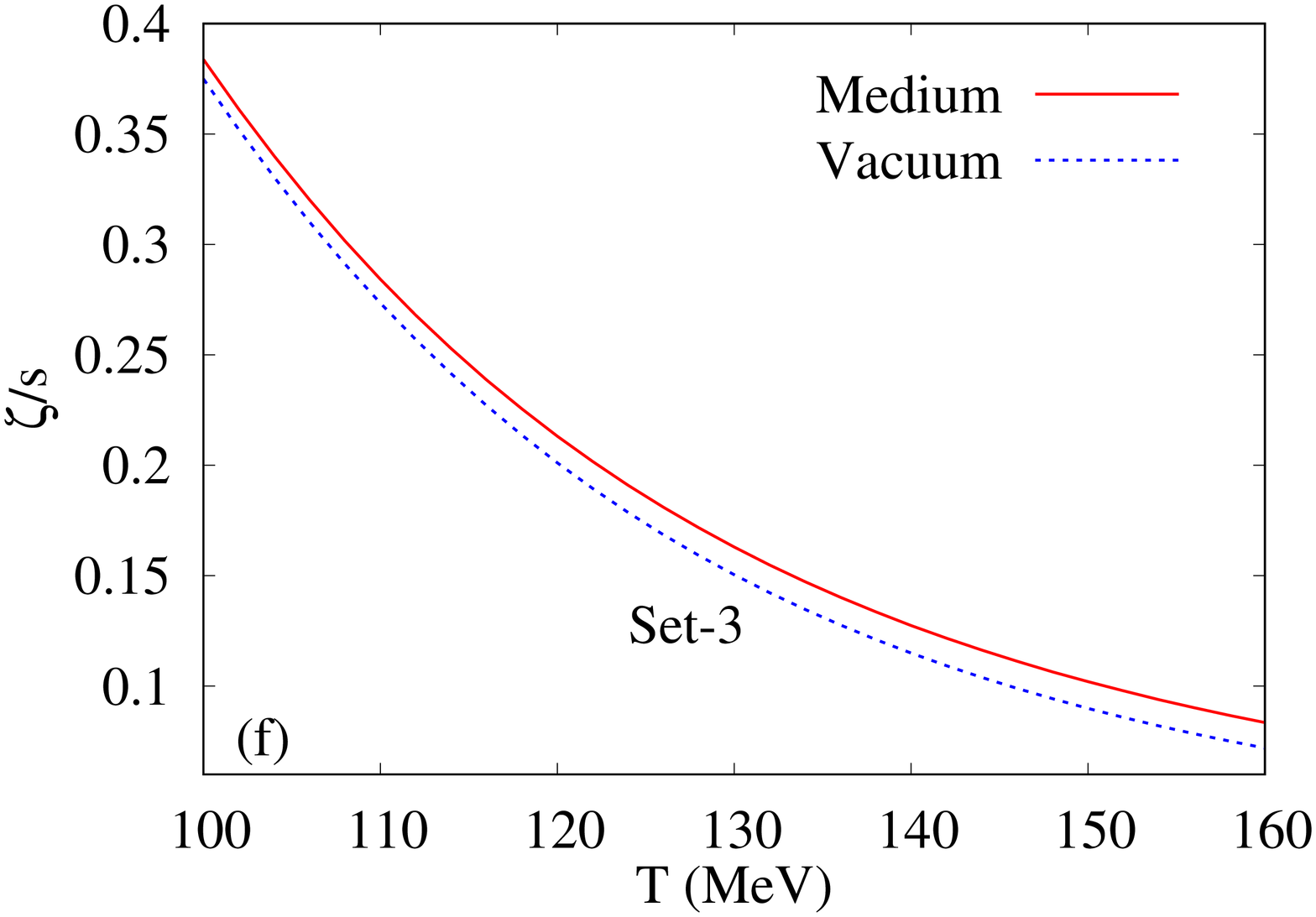}
	\end{center}
	\caption{Specific shear viscosity ($ \eta/s $) and specific bulk viscosity ($ \zeta/s $) as a function of temperature ($T$) for a pion-kaon-nucleon hadronic gas mixture for (a) Set-1, (b) Set-2 and (c) Set-3 of chemical potentials of individual components with and without including medium effects.}
	\label{fig:etazetas}
\end{figure}

For $\pi\pi$, we consider the $\sigma$ and $\rho$ exchange, whereas for $\pi K$ and $\pi N$ we consider $K^*$ and $\Delta$ exchange. The decay width of $\sigma$ and $\rho$ are taken from Refs.~\cite{Wiranata:2013oaa} and they are given by
\begin{eqnarray}
\Gamma_\sigma(s) &=& 2.06 q~, \\
\Gamma_\rho(s) &=& 0.095 q \FB{\frac{q/m_\pi}{1+q^2/m_\rho^2}}^2
\end{eqnarray}
where, $q=\frac{1}{2}\sqrt{s-4m_\pi^2}$. The decay widths of $K^*$ and $\Delta$ are calculated from the effective Lagrangians of Eqs.~\eqref{eq.Lagrangian.piK} and \eqref{eq.Lagrangian.piN} and are given by
\begin{eqnarray}
\Gamma_{K^*}(s) &=& \frac{g^2_{\pi KK^*}}{16\pi} F^2(s) \frac{1}{s^{3/2}} \lambda^{1/2}(s,m_\pi^2,m_K^2)
\TB{s-2m_K^2-2m_\pi^2+\frac{1}{s}(m_K^2-m_\pi^2)^2}, \\
\Gamma_{\Delta}(s) &=& \frac{f^2_{\pi N\Delta}}{192\pi m_\pi^2} F^2(s) \frac{1}{s^{5/2}} \lambda^{3/2}(s,m_\pi^2,m_N^2)
\TB{(\sqrt{s}-m_N)^2-m_\pi^2}
\end{eqnarray}
where, the form factor $F(s) = \Lambda_{K,N}^2\TB{\Lambda_{K,N}^2+\frac{1}{4m_{K,N}^2}\lambda(s,m_\pi^2,m_{K,N}^2)}^{-1}$ with $\lambda(x,y,z)=x^2+y^2+z^2-2xy-2yz-2zx$ being the K\"all\'en function.

In Fig.~\ref{fig:phase}, the phase shifts in the different resonance channels ($\sigma, \rho, K^*$ and $\Delta$) are shown as a function of the center of mass energy. The slope of $\delta_\sigma$ is least among the other three due to the largest decay width of $\sigma$. 
All the $\delta_\rho$, $\delta_{K^*}$ and $\delta_\Delta$ show rapid changes around the respective pole mass values ($m_\rho, m_{K^*}$ and $m_\Delta$) of the center of mass energies. The $\delta_{K^*}$ is the steepest among all due to the smallest vacuum decay width of $K^*$ which is $\sim 50$ MeV.

Next in Figs.~\ref{fig:entropy}, the entropy density with interactions scaled by the cube of inverse temperature ($s/T^3$) has been depicted as a function of temperature for the three sets of chemical potentials. Shown with triangles is the result for a free gas of $\pi, K, N, \sigma, \rho, K^*$ and $\Delta$ for Set-1. They appear to be in good agreement~\cite{Wiranata:2013oaa,Venugopalan:1992hy}.
\begin{figure}[h]
	\begin{center}
		\includegraphics[angle=0, scale=0.55]{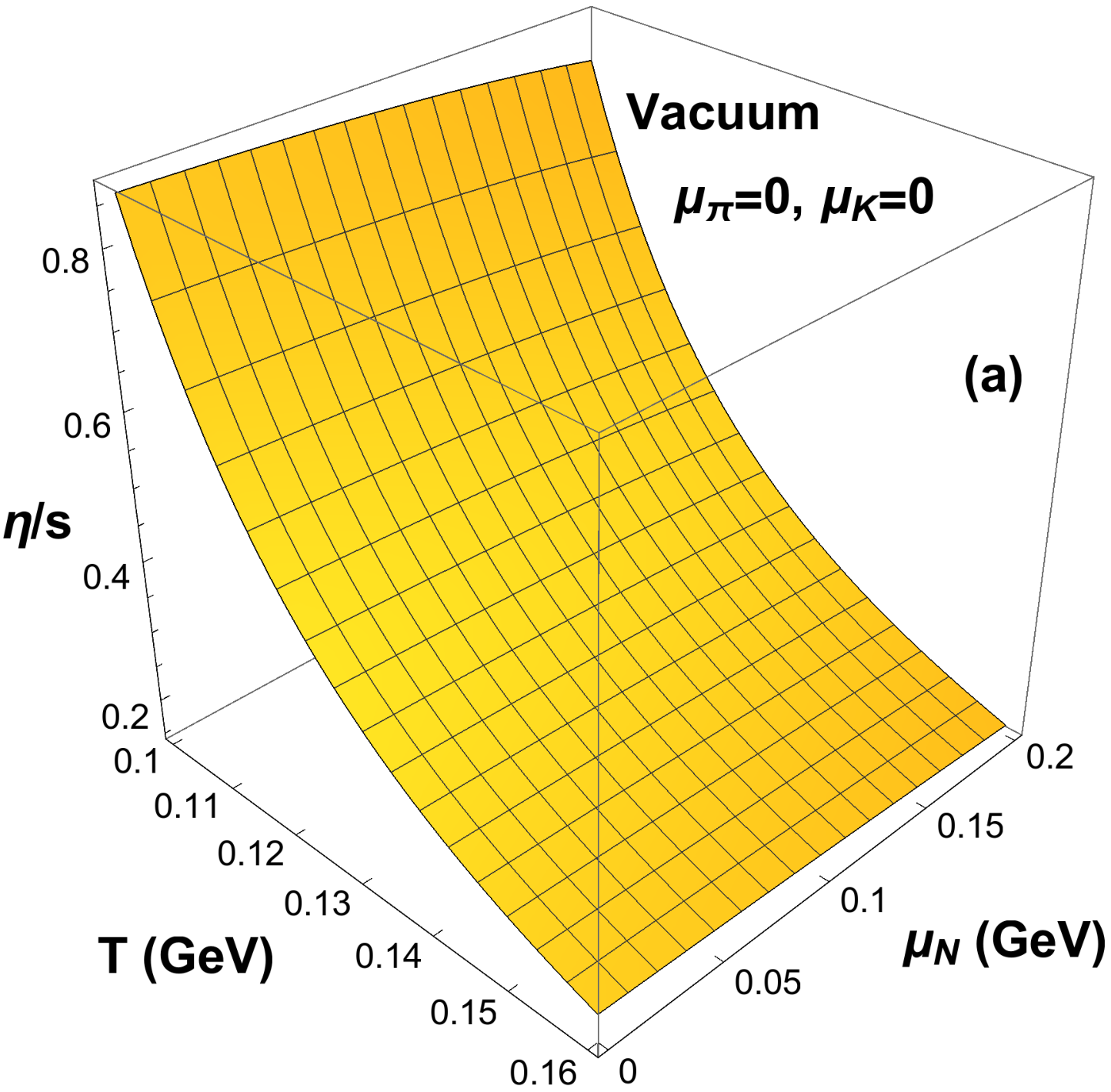}  ~~~~~ \includegraphics[angle=0, scale=0.55]{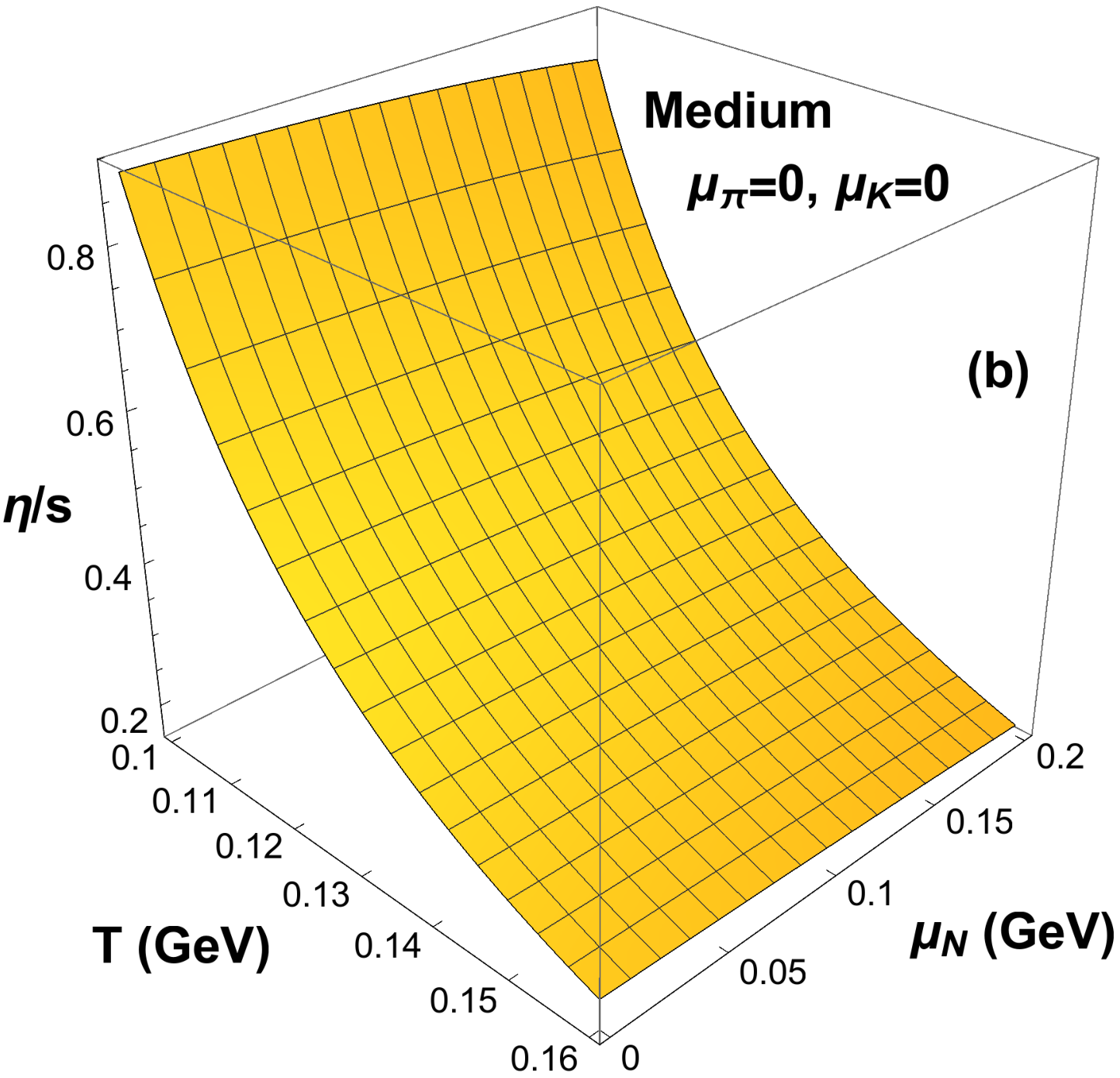}
	\end{center}
	\caption{ Shear viscosity to entropy density ratio ($\eta/s$) as a function of temperature and nucleon chemical potential at 
		$\mu_\pi = \mu_K = 0$ with (a) vacuum and (b) in-medium cross sections.}
	\label{Fig:3D.1}
\end{figure}
\begin{figure}[h]
	\begin{center}
		\includegraphics[angle=0, scale=0.55]{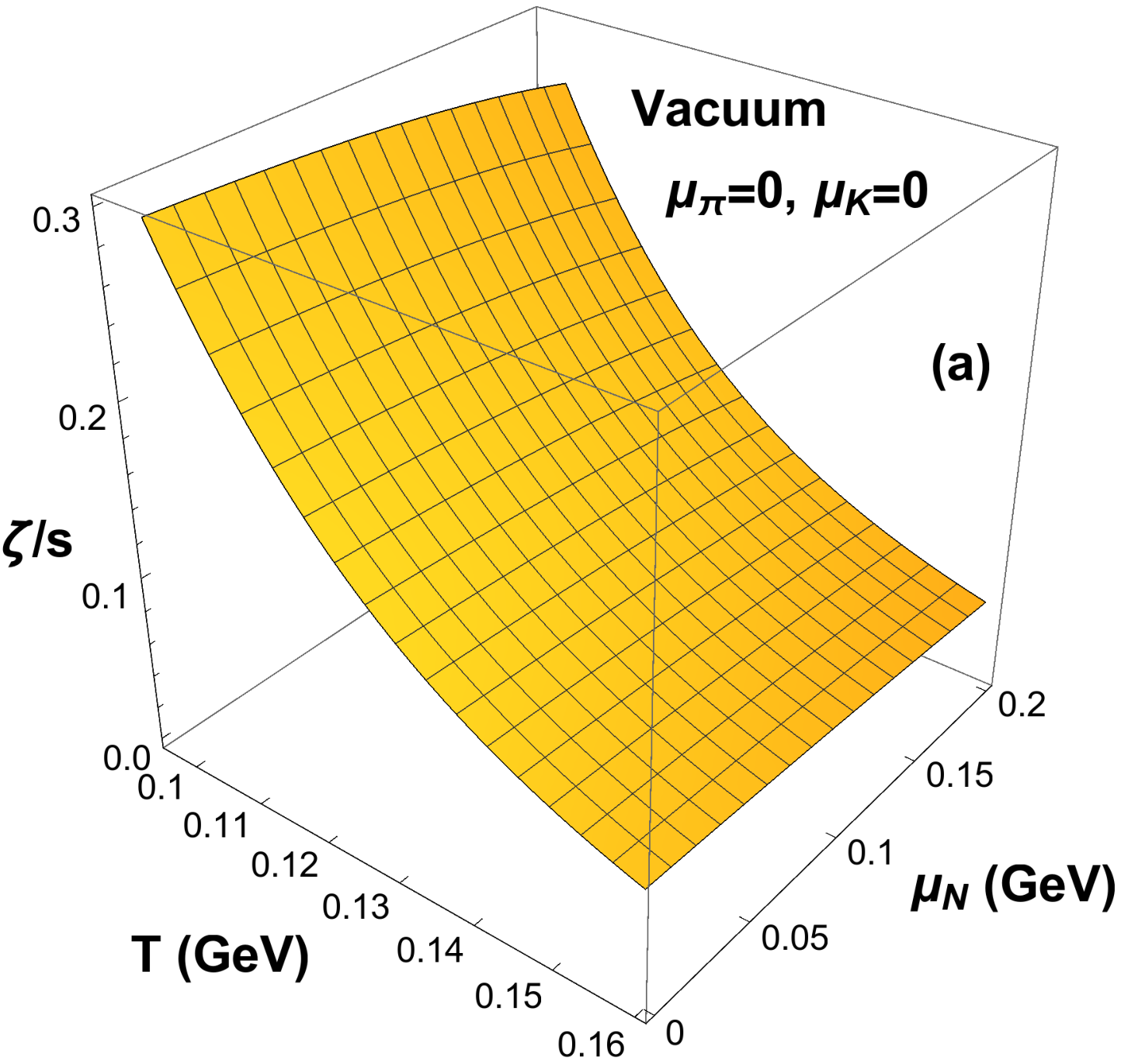} ~~~~~ \includegraphics[angle=0, scale=0.55]{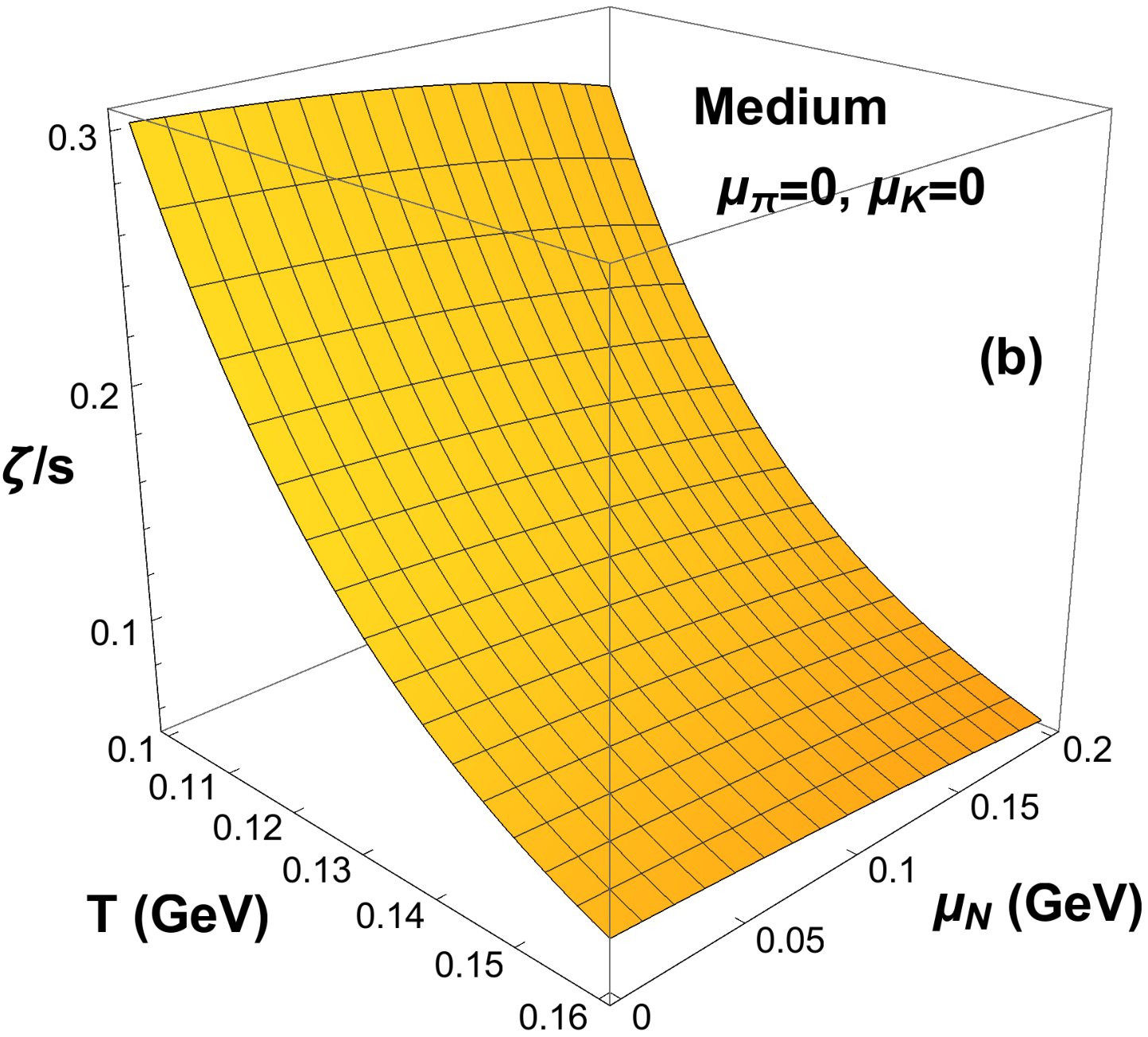}
	\end{center}
	\caption{ Bulk viscosity to entropy density ratio ($\zeta/s$) as a function of temperature and nucleon chemical potential at 
		$\mu_\pi = \mu_K = 0$ with (a) vacuum and (b) in-medium cross sections. }
	\label{Fig:3D.2}
\end{figure}
\begin{figure}[h]
	\begin{center}
		\includegraphics[angle=0, scale=0.55]{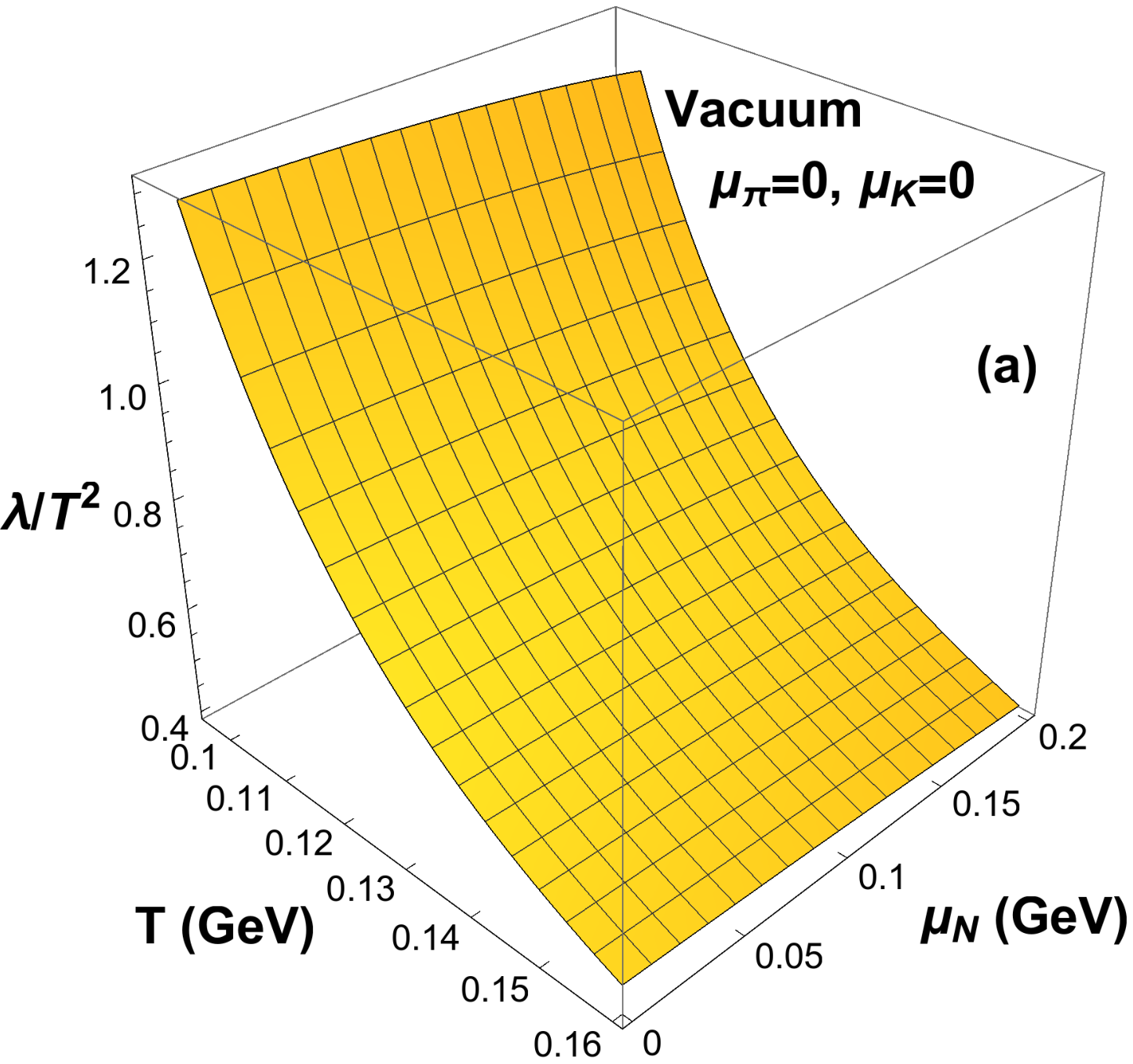} ~~~~~ \includegraphics[angle=0, scale=0.55]{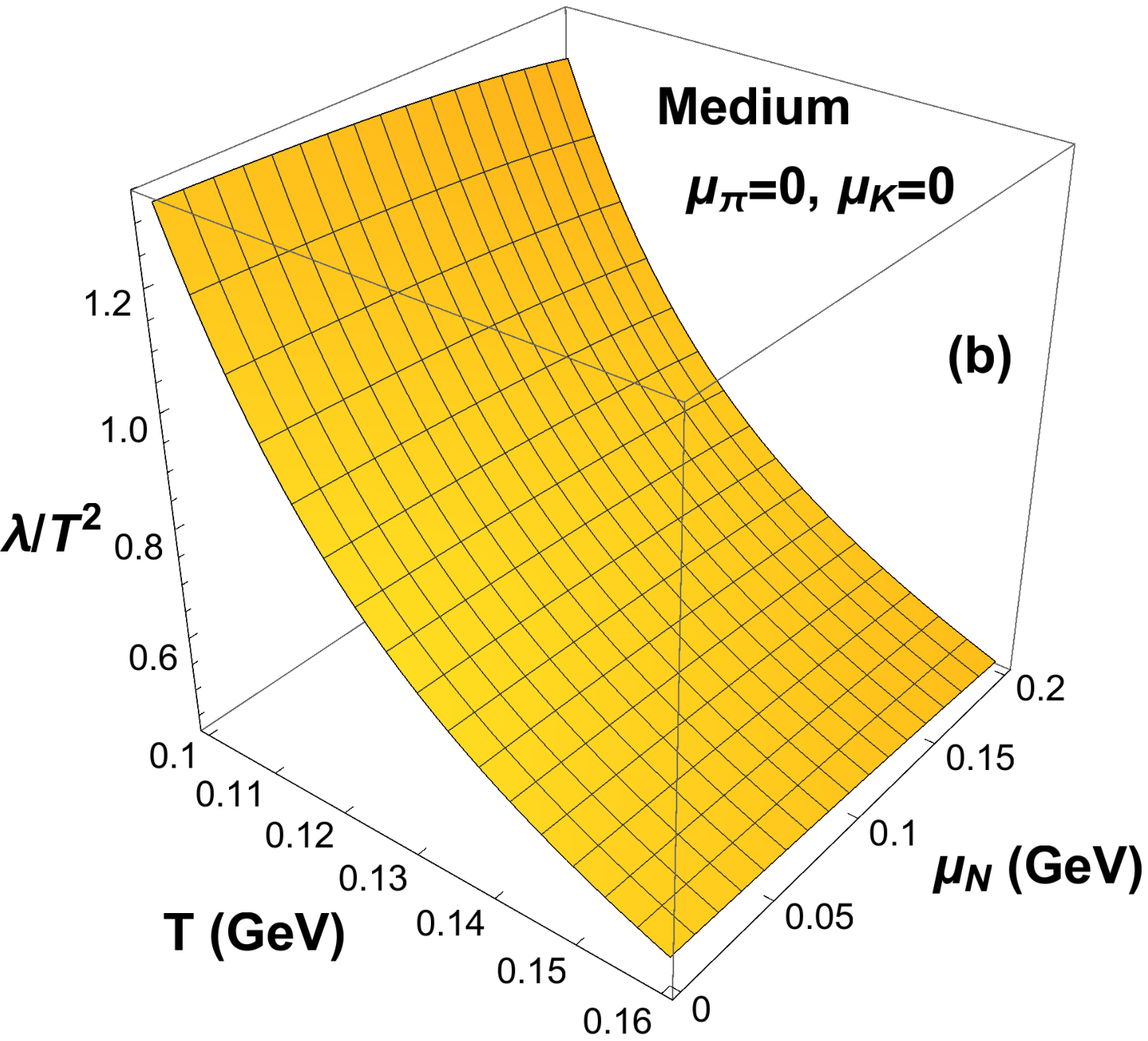}
	\end{center}
	\caption{ Thermal conductivity scaled with inverse of temperature squared ($\lambda/T^2$) as a function of 
		temperature and nucleon chemical potential at $\mu_\pi = \mu_K = 0$ with (a) vacuum and (b) in-medium cross sections.}
	\label{Fig:3D.3}
\end{figure}
\begin{figure}[h]
	\begin{center}
		\includegraphics[angle=-90, scale=0.35]{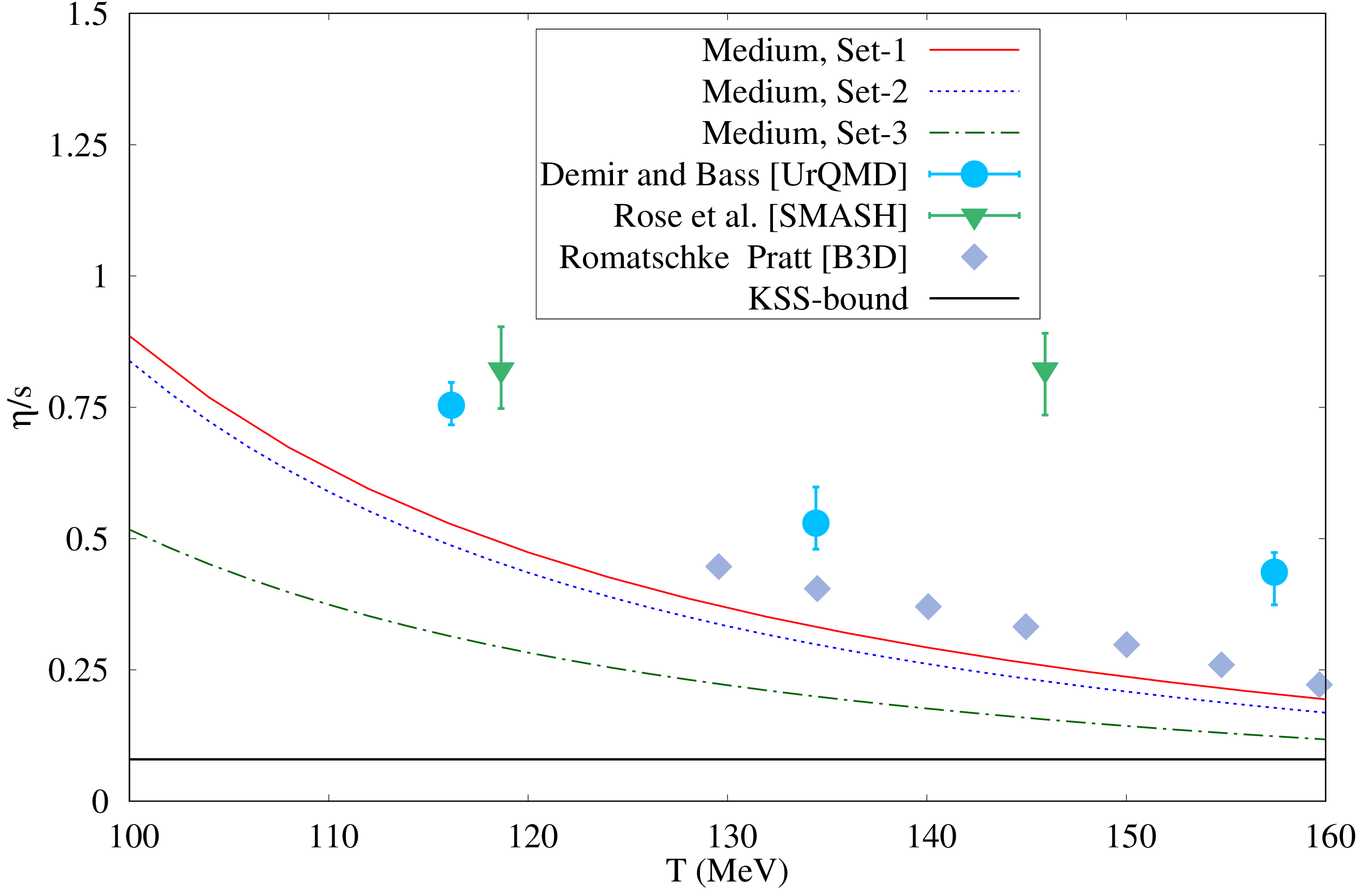}  
	\end{center}
	\caption{The result obtained in this paper compared to various data of the specific shear viscosity $\eta/s$ as a function of temperature
		available in the literature~\cite{Demir,Romas,Rose}. A line of KSS bound has been drawn as a reference. }
	\label{Fig_etabysexp_vs_T}
\end{figure}

We are now in a position to plot the specific viscosities.
In Figs.~\ref{fig:etazetas}(d)-(f) $\eta/s$ is plotted with temperature for three sets of chemical potentials. We see a monotonic decrease of the specific shear viscosity with increase of temperature. This is because of the fact that the entropy density rises with increase of temperature with a rate faster than $\eta$. The increase in magnitude of $\eta$ for different sets of increasing chemical potential, even though there is a decrease in relaxation time with the increase in chemical potential, is mainly due to the increase in density which turns out to be the governing factor here. $\eta/s$ on the other hand decreases with the increase in chemical potential because with the increase in chemical potential the entropy increases faster due to the rapid increase in the degrees of freedom, and the value of $\eta/s$ is within the KSS-bound. Here we note that with the inclusion of more resonances the shear viscosity decreases as one approaches temperatures close to the critical temperature. Owing to the corresponding increase in the entropy density there is a further decrease in the value of $\eta/s$~\cite{Wiranata:2013oaa}. 

We now  show the results of specific bulk viscosity. In Figs.~\ref{fig:etazetas}(d)-(f) one can also observe the sizeable difference in magnitude of 
$\zeta/s$  calculated for vacuum and medium. The figures show increase in magnitude of $\zeta$ for increasing value of chemical potential just like in the case for $\eta$, which is again attributed to the dominance of increase in density over the decrease in magnitude of relaxation time. And like $\eta/s$ we find that $\zeta/s$ increases with increase in chemical potential.

In Figs.~\ref{Fig:3D.1} and \ref{Fig:3D.2}, we have studied the effect of temperature and baryonic chemical potential on $\eta/s$ and $\zeta/s$ with and without medium effects taken into consideration, for $\mu_{\pi}=\mu_{K}=0$. We find that on introducing the medium effects, there is an increase in the magnitude but very little change in its behaviour. Also, with increase in baryonic chemical potential the value of $\eta /s$ decreases while the value increases for $\zeta /s$. The drop in the value of $\eta/s$ with increase in temperature is found to be sharper at lower baryonic chemical potential while we observe the opposite trend in the case of $\zeta/s$ where the drop is sharper at higher chemical potential. At lower temperature the effect of the baryonic chemical potential on $\zeta/s$ is found to be more pronounced than at higher temperature, this pattern is also ovserved in the case of $\eta/s$. 
The analogous plots for the variation of $\lambda/T^2$ as a function of $T$ and $\mu_N$ shown in Fig.~\ref{Fig:3D.3}(a) and (b) show features similar to that of $\eta/s$.

Finally, to check consistency we have plotted in Fig.~\ref{Fig_etabysexp_vs_T}, $ \eta/s $ calculated using different techniques such as hadronic cascades URQMD~\cite{Demir}, B3D~\cite{Romas} and SMASH~\cite{Rose} along with our result using medium dependent cross-sections for the three sets of chemical potentials.  In particular, the in-medium $\eta/s$ calculated with Set-1 i.e. for $\mu_\pi=0$, $\mu_K=0$ and $\mu_N=0$ are in reasonable agreement with that of~\cite{Romas}.

\section{SUMMARY}
In this work we have considered a hot and dense hadronic gas mixture consisting of pions, kaons and nucleons 
which are the most important components of the system produced during the later stages of heavy ion collisions. 
We have endeavored  to present a systematic study of the relaxation times, viscous coefficients and thermal conductivity for a  system consisting only of pions, a system of pions and kaons and finally for a pion-kaon-nucleon system using the Boltzmann transport equation which has been linearised using the Enskog expansion. The key ingredient is the use of in-medium cross-sections which were obtained using one-loop corrected thermal propagators in the matrix elements for $\pi\pi$, $\pi K$ and $\pi N$ scattering. The suppression of the in-medium cross-sections at finite temperature and density is reflected in the enhancement of relaxation times. This in turn results in observable modification of the temperature dependence of $\eta$, $\zeta$ and $\lambda/T^2$. However, the temperature dependence of $\eta/s$ and $\zeta/s$, where the entropy density $s$ also contains the effect of interactions, is much less affected by the medium. On comparison, the value of $\eta/s$ in the medium for vanishing chemical potentials is found to be in agreement with existing estimates in the literature. 

\section*{Acknowledgments}
P.K., S.G. and U.G. acknowledge the hospitality of Variable Energy Cyclotron Centre, Kolkata where most of the work was done. They also acknowledge Department of Atomic Energy, Government of India for providing financial support.

\appendix

\section{THERMODYNAMIC QUANTITIES} \label{sec.app.a}
The thermodynamic quantities like energy density, number density, pressure and enthalpy of the three component system  consisting of pions, kaons and nucleons can be expressed in terms of the sum of series of Bessels function as $S_n^\alpha (z_\pi)$, $R_n^\alpha (z_K)$ and $T_n^\alpha (z_N)$, where $z_\pi=m_\pi /T$, $z_K=m_K /T$ and $z_N=m_N /T$. These quantities are given as:
\begin{eqnarray}
n_\pi &=& g_\pi \int \frac{d^3p_\pi}{(2\pi)^3} f_\pi^{(0)} (p_\pi)= \FB{\frac{g_\pi}{2 \pi^2}} z_\pi ^2 T^3 S_2^1 (z_\pi)~, \\
P_\pi &=& g_\pi \int \frac{d^3p_\pi}{(2\pi)^3} \frac{\vec{p_\pi}^2}{3E_{p_\pi}} f_\pi^{(0)} (p_\pi)= \FB{\frac{g_\pi}{2 \pi^2}} z_\pi ^2 T^4 S_2^2 (z_\pi)~, \\
n_\pi e_\pi &=& g_\pi \int \frac{d^3p_\pi}{(2\pi)^3} E_{p_\pi} f_\pi ^{(0)}(p_\pi)= \FB{\frac{g_\pi}{2 \pi^2}} z_\pi ^2 T^4 \TB{z_\pi S_3^1(z_\pi)-S_2^2(z_\pi)}, \\
n_\pi h_\pi &=& n_\pi z_\pi \frac{S_3^1(z_\pi)}{S_2^1(z_\pi)}
\end{eqnarray}
where $E_{p_\pi}=\sqrt{\vec{p}_\pi^2+m_\pi^2}$ and $f_\pi^{(0)}(p_\pi)=[e^{\beta(E_{p_\pi}-\mu_\pi)}-1]^{-1}$. 
Making use of the formula 
\begin{eqnarray}
[a-1]^{-1}=\sum_{n=1}^{\infty}~(a^{-1})^n~~~\text{for}~~ \MB{a}<1~,
\end{eqnarray}
the distribution function can be expanded, so that the three momentum integrals in the above equations could be analytically performed and expressed in terms of the following infinite series
\begin{eqnarray}
S_n^\alpha(z_\pi)=\sum_{k=1}^{\infty}~e^{{k\mu_\pi}/T}~k^{-\alpha}~K_n(kz_\pi)
\end{eqnarray}
where $K_n(x)$ is the modified Bessel function of order $n$ whose integral representation is 
\begin{equation}
K_n(x)=\frac{2^n n!}{(2n)!~x^n}\int_{x}^{\infty}d\tau(\tau^2-x^2)^{n-\frac{1}{2}}e^{-\tau}
\end{equation}
or
\begin{equation}
K_n(x)=\frac{2^n n! (2n-1)}{(2n)!x^n}\int_{x}^{\infty}\tau~ d\tau (\tau^2-x^2)^{n-\frac{3}{2}}~e^{-\tau}.
\end{equation}
The expression for thermodynamic quantities mentioned above will be similar for kaons and nucleons except the term $S_n^\alpha(z_\pi)$ will be replaced by $R_n^\alpha(z_K)$ for kaons and $T_n^\alpha(z_N)$ for nucleons where
\begin{eqnarray}
R_n^\alpha(z_K)=\sum_{k=1}^{\infty}~e^{{k\mu_K}/T}~k^{-\alpha}~K_n(kz_K)
\end{eqnarray}
 and 
 \begin{eqnarray}
T_n^\alpha(z_N)=\sum_{k=1}^{\infty}~(-1)^{k-1}~e^{{k\mu_N}/T}~k^{-\alpha}~K_n(kz_N)~.
 \end{eqnarray}


\section{USEFUL EXPRESSIONS} \label{sec.app.b}
The transport equation for each species is given by
\begin{equation}
p^\mu\partial_\mu f_k^{(0)}(x,p)=-\frac{\delta f(x,p)}{\tau_k} E_k
\end{equation}
where on the right hand side of the equation, we have made use of relaxation time approximation. The time and space derivatives 
(in the local rest frame) present in the left hand side of the above equation will be replaced by the derivatives of the thermodynamics parameters. The equation then reduces to 
\begin{eqnarray}
(p_k\cdot u)\left[\frac{p_k\cdot u}{T^2}DT + D\left(\frac{\mu_k}{T}\right)~-~\frac{p_k^\mu}{T}Du_\mu \right] +
p^\mu\left[\frac{p_k\cdot u}{T^2} \nabla_\mu T + \nabla_\mu \left(\frac{\mu_k}{T}\right)-\frac{p_k^\nu}{T}\nabla_\mu u_\nu \right]=-\frac{\delta f(x,p)}{\tau_k} E_k~.
\end{eqnarray}
The conservation equations
\begin{eqnarray}
\partial_\mu N_k^\mu=0,~~~ Dn_k=-n_k \partial_\mu u^\mu ~~\text{and}~~ \sum_{k} n_k De_k=-\sum_{k}P_k\partial_\mu u^\mu 
\end{eqnarray}
with $N^\mu=n U^\mu$ and total $P=p_\pi+p_K+p_N$ can be expanded in terms of the derivative with respect to temperature and chemical potential over temperature as
\begin{eqnarray}
\frac{\partial n_\pi}{\partial T} DT + \frac{\partial n_\pi}{\partial(\mu_\pi/T)}D\left(\frac{\mu_\pi}{T}\right) + \frac{\partial n_K}{\partial(\mu_K/T)}D\left(\frac{\mu_K}{T}\right)+
\frac{\partial n_N}{\partial(\mu_N/T)}D\left(\frac{\mu_N}{T}\right)&=&-n_\pi~\partial_\mu u^\mu~, \\
\frac{\partial n_K}{\partial T} DT + \frac{\partial n_\pi}{\partial(\mu_\pi/T)}D\left(\frac{\mu_\pi}{T}\right)~+~\frac{\partial n_K}{\partial(\mu_K/T)}D\left(\frac{\mu_K}{T}\right) + \frac{\partial n_N}{\partial(\mu_N/T)}D\left(\frac{\mu_N}{T}\right)&=&-n_K~\partial_\mu u^\mu~,\\
\frac{\partial n_N}{\partial T} DT+\frac{\partial n_\pi}{\partial(\mu_\pi/T)}D\left(\frac{\mu_\pi}{T}\right)+\frac{\partial n_K}{\partial(\mu_K/T)}D\left(\frac{\mu_K}{T}\right)+ \frac{\partial n_N}{\partial(\mu_N/T)}D\left(\frac{\mu_N}{T}\right)&=&-n_N~\partial_\mu u^\mu~, \\
\left[n_\pi\frac{\partial e_\pi}{\partial T}+n_K\frac{\partial e_K}{\partial T}+n_N\frac{\partial e_N}{\partial T} \right]+n_\pi\frac{\partial e_\pi}{\partial(\mu_\pi/T)}  D\left(\frac{\mu_\pi}{T}\right) && \nn \\
 + n_K\frac{\partial e_K}{\partial(\mu_K/T)}D\left(\frac{\mu_K}{T}\right)+n_N\frac{\partial e_N}{\partial(\mu_N/T)}D\left(\frac{\mu_N}{T}\right)&=&-P\partial_\mu u^\mu~. \nonumber 
\end{eqnarray}
Making use of the expressions obtained in Appendix~\ref{sec.app.a} in the above equations and then solving for $DT$, $D\left( \frac{\mu_\pi}{T}\right)$, $D\left( \frac{\mu_K}{T}\right)$ and $D\left( \frac{\mu_N}{T}\right)$ we get
\begin{eqnarray}
DT &=& T~(1-\gamma')~\partial_\mu u^\mu~, \\
TD\left(\frac{\mu_\pi}{T}\right)&=& [(\gamma_\pi''-1)-T\gamma_\pi''']~\partial_\mu u^\mu~, \\
TD\left(\frac{\mu_K}{T}\right)&=& [(\gamma_K''-1)-T\gamma_K''']~\partial_\mu u^\mu~, \\
TD\left(\frac{\mu_N}{T}\right)&=& [(\gamma_N''-1)-T\gamma_N''']~\partial_\mu u^\mu
\end{eqnarray}
where
\begin{eqnarray}
\gamma' &=& \frac{1}{X}\Bigg[g_\pi\SB{z_\pi^3\FB{4R_2^0S_2^0T_2^0S_3^1+R_2^0T_2^0S_3^0S_2^1\frac{}{}}
	+z_\pi^4\FB{R_2^0T_2^0(S_2^0)^2-R_2^0T_2^0(S_3^0  )^2\frac{}{}}} \nonumber \\
 && +g_K\SB{z_K^3\FB{4R_2^0S_2^0T_2^0R_3^1+S_2^0T_2^0R_3^0R_2^1\frac{}{}}
 	+z_K^4\FB{S_2^0T_2^0(R_2^0)^2-S_2^0T_2^0(R_3^0)^2 \frac{}{}}} \nonumber \\
 && +g_N\SB{z_N^3\FB{4R_2^0S_2^0T_2^0T_3^1+R_2^0S_2^0T_3^0T_2^1\frac{}{}}
 	+z_N^4\FB{R_2^0S_2^0(T_2^0)^2-R_2^0S_2^0(T_3^0)^2\frac{}{}}}\Bigg], 
\end{eqnarray}
\begin{eqnarray}
\gamma_\pi''&=&\frac{1}{X}\Bigg[g_\pi\SB{-5z_\pi^2R_2^0T_2^0(S_2^1)^2+z_\pi^3\FB{3R_2^0S_2^0T_2^0S_3^1+3R_2^0T_2^0S_3^0S_2^1\frac{}{}}
+z_\pi^4\FB{R_2^0T_2^0(S_2^0)^2-R_2^0T_2^0(S_3^0)^2\frac{}{}}} \nonumber \\
&& +g_K\SB{-z_K^2S_2^0T_2^0(R_2^1)^2 +z_K^3\FB{3R_2^0S_2^0T_2^0R_3^1+2R_3^0S_2^0T_2^0R_2^1\frac{}{}}
+z_K^4\FB{S_2^0T_2^0(R_2^0)^2-S_2^0T_2^0(R_3^0)^2\frac{}{}}} \nonumber\\
&& +g_N\SB{-z_N^2R_2^0S_2^0(T_2^1)^2+z_N^3\FB{3R_2^0S_2^0T_2^0T_3^1+2R_2^0S_2^0T_3^0T_2^1\frac{}{}}
+z_N^4\FB{R_2^0S_2^0(T_2^0)^2-R_2^0S_2^0(T_3^0)^2\frac{}{}}}\Bigg],
\end{eqnarray}
\begin{eqnarray}
\gamma_\pi'''&=&\frac{1}{X}\Bigg[g_\pi\SB{z_\pi^4R_2^0T_2^0S_2^0S_2^1\frac{}{}}
+g_K\SB{z_K^3\FB{4R_2^0T_2^0S_2^1R_3^1+T_2^0R_3^0R_2^1S_2^1\frac{}{}}
+z_K^4\FB{T_2^0S_2^1(R_2^0)^2-T_2^0S_2^1(R_3^0)^2 \frac{}{}} \right. \nonumber \\
&& \left. -z_\pi z_K^2T_2^0S_3^0(R_2^1)^2 
+z_\pi z_K^3\FB{R_3^0S_3^0T_2^0R_2^1-R_2^0S_3^0T_2^0R_3^1\frac{}{}} }
+g_N\SB{z_N^3\FB{4R_2^0T_2^0S_2^1T_3^1+R_2^0T_3^0S_2^1T_2^1\frac{}{}} \right. \nonumber\\ 
&& \left. +z_N^4\FB{R_2^0S_2^1(T_2^0)^2-R_2^0S_2^1(T_3^0)^2\frac{}{}}-z_\pi z_N^2R_2^0S_3^0(T_2^1)^2
+z_\pi z_N^3\FB{R_2^0S_3^0T_3^0T_2^1-R_2^0S_3^0T_2^0T_3^1\frac{}{}} }\Bigg] ,
\end{eqnarray}
\begin{eqnarray}
\gamma_K''&=&\frac{1}{X}\Bigg[g_\pi\SB{-z_\pi^2R_2^0T_2^0(S_2^1)^2+z_\pi^3\FB{3R_2^0S_2^0T_2^0S_3^1+2R_2^0T_2^0S_3^0S_2^1\frac{}{}}
+z_\pi^4\FB{R_2^0T_2^0(S_2^0)^2-R_2^0T_2^0(S_3^0)^2\frac{}{}}} \nonumber \\
&&+g_K\SB{-5z_K^2S_2^0T_2^0(R_2^1)^2 +z_K^3\FB{3R_2^0S_2^0T_2^0R_3^1+3R_3^0S_2^0T_2^0R_2^1\frac{}{}}
+z_K^4\FB{S_2^0T_2^0(R_2^0)^2-S_2^0T_2^0(R_3^0)^2\frac{}{}}} \nonumber\\
&&+g_N\SB{-z_N^2R_2^0S_2^0(T_2^1)^2+z_N^3\FB{3R_2^0S_2^0T_2^0T_3^1+2R_2^0S_2^0T_3^0T_2^1\frac{}{}}
+z_N^4\FB{R_2^0S_2^0(T_2^0)^2-R_2^0S_2^0(T_3^0)^2\frac{}{}}}\Bigg] ,
\end{eqnarray}
\begin{eqnarray}
\gamma_K'''&=&\frac{1}{X}\Bigg[g_\pi\SB{z_\pi^3\FB{4T_2^0S_2^0R_2^1S_3^1+T_2^0S_3^0R_2^1S_2^1\frac{}{}}+z_\pi^4\FB{T_2^0R_2^1(S_2^0)^2
-T_2^0R_2^1(S_3^0)^2\frac{}{}}-z_Kz_\pi^2T_2^0R_3^0(S_2^1)^2 \right. \nonumber \\
&&\left. +z_Kz_\pi^3\FB{T_2^0R_3^0S_3^0S_2^1 - T_2^0R_3^0S_2^0S_3^1\frac{}{}}}+g_K\SB{z_k^4S_2^0T_2^0R_2^0R_2^1\frac{}{}}
+g_N\SB{z_N^3\FB{4T_2^0S_2^0R_2^1T_3^1+S_2^0T_3^0R_2^1T_2^1\frac{}{}} \right. \nonumber\\ 
&&\left. +z_N^4\FB{S_2^0R_2^1(T_2^0)^2-S_2^0R_2^1(T_3^0)^2\frac{}{}}-z_K z_N^2S_2^0R_3^0(T_2^1)^2
+z_Kz_N^3\FB{S_2^0R_3^0T_3^0T_2^1-S_2^0R_3^0T_2^0T_3^1\frac{}{}}}\Bigg] ,
\end{eqnarray}
\begin{eqnarray}
\gamma_N''&=&\frac{1}{X}\Bigg[g_\pi\SB{-z_\pi^2R_2^0T_2^0(S_2^1)^2+z_\pi^3\FB{3R_2^0S_2^0T_2^0S_3^1+2R_2^0T_2^0S_3^0S_2^1\frac{}{}}
+z_\pi^4\FB{R_2^0T_2^0(S_2^0)^2-R_2^0T_2^0(S_3^0)^2\frac{}{}}} \nonumber \\
&&+g_K\SB{-z_K^2S_2^0T_2^0(R_2^1)^2 +z_K^3\FB{3R_2^0S_2^0T_2^0R_3^1+2R_3^0S_2^0T_2^0R_2^1\frac{}{}}
+z_K^4\FB{S_2^0T_2^0(R_2^0)^2-S_2^0T_2^0(R_3^0)^2\frac{}{}}} \nonumber\\
&&+g_N\SB{-5z_N^2R_2^0S_2^0(T_2^1)^2+z_N^3\FB{3R_2^0S_2^0T_2^0T_3^1+3R_2^0S_2^0T_3^0T_2^1\frac{}{}}
+z_N^4\FB{R_2^0S_2^0(T_2^0)^2-R_2^0S_2^0(T_3^0)^2\frac{}{}}}\Bigg] ,
\end{eqnarray}
\begin{eqnarray}
\gamma_N'''&=&\frac{1}{X}\Bigg[g_\pi\SB{z_\pi^3\FB{4S_2^0R_2^0T_2^1S_3^1+R_2^0S_3^0T_2^1S_2^1\frac{}{}}+z_\pi^4\FB{R_2^0T_2^1(S_2^0)^2
-R_2^0T_2^1(S_3^0)^2\frac{}{}}-z_Nz_\pi^2R_2^0T_3^0(S_2^1)^2 \right. \nonumber\\
&&\left. +z_Nz_\pi^3\FB{R_2^0S_3^0T_3^0S_2^1 - R_2^0S_2^0T_3^0S_3^1\frac{}{}}} +g_N\SB{z_N^4S_2^0T_2^0R_2^0T_2^1\frac{}{}}
+g_K\SB{z_K^3\FB{4R_2^0S_2^0T_2^1R_3^1+R_3^0S_2^0R_2^1T_2^1\frac{}{}} \right. \nonumber\\
&&\left. +z_K^4\FB{(R_2^0)^2S_2^0T_2^1 -(R_3^0)^2S_2^0T_2^1\frac{}{}}-z_Nz_K^2S_2^0T_3^0(R_2^1)^2
+z_Nz_K^3\FB{S_2^0R_3^0T_3^0R_2^1-R_2^0S_2^0T_3^0R_3^1\frac{}{}}}\Bigg] ,
\end{eqnarray}
and the term $X$ appearing in the above expressions of $\gamma$'s is given by
\begin{eqnarray}
X&=&g_\pi\TB{-z_\pi^2R_2^0T_2^0(S_2^1)^2+z_\pi^3\FB{3R_2^0S_2^0T_2^0S_3^1+2R_2^0T_2^0S_3^0S_2^1\frac{}{}}
+z_\pi^4\FB{R_2^0T_2^0(S_2^0)^2-R_2^0T_2^0(S_3^0)^2\frac{}{}}} \nonumber \\
&&+g_K\TB{-z_K^2S_2^0T_2^0(R_2^1)^2 +z_K^3\FB{3R_2^0S_2^0T_2^0R_3^1+2R_3^0S_2^0T_2^0R_2^1\frac{}{}}
+z_K^4\FB{S_2^0T_2^0(R_2^0)^2-S_2^0T_2^0(R_3^0)^2\frac{}{}}} \nonumber\\
&&+g_N\TB{-z_N^2R_2^0S_2^0(T_2^1)^2+z_N^3\FB{3R_2^0S_2^0T_2^0T_3^1+2R_2^0S_2^0T_3^0T_2^1\frac{}{}}
+z_N^4\FB{R_2^0S_2^0(T_2^0)^2-R_2^0S_2^0(T_3^0)^2\frac{}{}}  }~.
\end{eqnarray}

\bibliographystyle{apsrev4-1}
\bibliography{pallavi}

\end{document}